\documentclass[fleqn,usenatbib]{mnras}

\usepackage{newtxtext,newtxmath}
\usepackage{mathpazo}

\usepackage[T1]{fontenc}

\DeclareRobustCommand{\VAN}[3]{#2}
\let\VANthebibliography\thebibliography
\def\thebibliography{\DeclareRobustCommand{\VAN}[3]{##3}\VANthebibliography}

\usepackage{graphicx}	
\usepackage{amsmath}	

\title[Reconstructing SN gravitational-wave waveforms]{Waveform Reconstruction of Core-Collapse Supernova Gravitational
Waves with Improved Multisynchrosqueezing Transform 
}

\author[Y. Yuan et al.]{
Yong Yuan $^{1}$,
Ao-Ran Wang$^{2}$,
Zhuo-Tao Li$^{3}$,
Gang Yu $^{2}$,
Hou-Jun L\"{u}$^{4}$,
Peng Xu$^{1}$ \thanks{E-mail:  xupeng@imech.ac.cn},
Xi-Long Fan $^{5}$ \thanks{Corresponding author:  xilong.fan@whu.edu.cn}
\\
$^{1}$Center for Gravitational Wave Experiment, National Microgravity Laboratory, Institute of Mechanics, Chinese Academy of Sciences, Beijing, China\\
$^{2}$School of Automation and Electrical Engineering, University of Jinan, Jinan, 250000, China; \\
$^{3}$ SUPA, University of Glasgow, Glasgow, G12 8QQ, United Kingdom\\
$^{4}$ Guangxi Key Laboratory for Relativistic Astrophysics, School of Physical Science and Technology, Guangxi University, Nanning, Guangxi, China\\
$^{5}$School of Physics Science And Technology, Wuhan University, No.299 Bayi Road, Wuhan, Hubei, China\\
}

\date{Accepted XXX. Received YYY; in original form ZZZ}

\pubyear{\the\year{}}

\begin{document}
\label{firstpage}
\pagerange{\pageref{firstpage}--\pageref{lastpage}}
\maketitle

\begin{abstract}
Gravitational waves (GWs) from core-collapse supernovae (CCSNe) have been proposed as a means to probe the internal physical properties of supernovae. However, due to their complex time-frequency structure, effectively searching for and extracting GW signals from CCSNe remains an unsolved challenge. In this paper, we apply the improved multisynchrosqueezing transform (IMSST) method to reconstruct simulated GW data based on the advanced LIGO (aLIGO) and Einstein Telescope (ET) detectors. These data are generated by the magnetorotational and neutrino-driven mechanisms, and we use the match score as the criterion for evaluating the quality of the reconstruction. To assess whether the reconstructed waveforms correspond to true GW signals, we calculate the false alarm probability of reconstruction (FAPR). For GW sources located at 10 kpc and datasets where the waveform amplitudes are normalized to $5 \times 10^{-21}$ observed by aLIGO,  FAPR are $2.1 \times 10^{-2}$ and $6.2 \times 10^{-3}$, respectively.  For GW sources at 100 kpc and with waveform amplitudes normalized to $5 \times 10^{-21}$ observed by ET, FAPR are $1.3 \times 10^{-1}$ and $1.5 \times 10^{-2}$, respectively. When the gravitational wave strain reaches $7 \times 10^{-21}$ and the match score threshold is set to 0.75, the IMSST method achieves maximum reconstruction distances of approximately 37 kpc and 317 kpc for aLIGO and ET, respectively. Finally, we compared the performance of IMSST and STFT in waveform reconstruction based on the ET. The results show that the maximum reconstructable distance using STFT is 186 kpc. 
\end{abstract}




\section{Introduction}

In 2015, advanced LIGO (aLIGO; \cite{LIGO_2015CQGra}) made the first-ever detection of gravitational waves (GWs) from the merger of two black holes, marking the beginning of the era of GW astronomy \citep{Abbott2016PhRvL}. With the addition of advanced Virgo (aVirgo; \cite{Acernese_2015}) to the network of ground-based GW detectors, the number of detected GW events steadily increased. By the end of the O3 phase, aLIGO and aVirgo had detected nearly 100 GW events originating from the mergers of compact binary systems \citep{Abbott2021GWTC3}. As KAGRA \citep{Abbott2022PTEP} joined the global network, alongside the continuous upgrades to ground-based GW detectors and the planned construction of the high frequency detector NEMO \citep{Ackley2020PASA}, the Einstein Telescope (ET) \citep{Punturo_2010CQGra} and Cosmic Explorer (CE) \citep{Abbott2017CQGra}, it is anticipated that an increasing number of GW events, as well as a wider variety of GW sources, will be detected \citep{Abbott2017CQGra, Maggiore2020JCAP}.

In recent decades, with the rapid advancement in multidimensional simulations of core-collapse supernovae (CCSNe), our understanding of the physical mechanisms behind GW emission in supernova explosions has deepened (see \cite{Janka_2012, Janka2016ARNPS, Muller2016PASA, Burrows2021Natur} for recent reviews). Through these simulations, it has been revealed that the frequency of GW generated by CCSNe also falls within the detection range of the aLIGO and aVirgo detectors \citep{Kotake2006RPPh, Moenchmeyer1991A&A, Muller2011hnse, Janka2007PhR}. GW leave the core around the time of collapse and can be used to directly probe the supernova engine, allowing us to understand the dynamics of moving matter \citep{Szczepanczyk2021PhRvD}. Unfortunately, previous analyses aimed at detecting GW signals from CCSNe have not yielded any significant detection candidates \citep{Abadie2012PhRvD, Abbott2016PhRvD, Abbott2017PhRvD, Abbott2019PhRvD, Abbott2020PhRvD, Abbott2021PhRvD, LIGO2024arXiv}.

A massive star, with an initial mass exceeding 8 $M_\odot$ at the zero-age main sequence, will reach the final stage of its life when its nuclear fuel is exhausted through nuclear reactions. During this phase, core collapse is anticipated if the mass of the core surpassed the effective Chandrasekhar mass limit \citep{Baron_apj1990, Bethe_1990}. There are several theories regarding the mechanisms by which CCSNe generate GW, but the two most popular theories are the neutrino-driven mechanism \citep{Janka_2012, Bethe_1990, Bethe_apj1985} and the magnetorotational mechanism \citep{Scheidegger_2008, Janka_2012, Kotake_2012PTEP, Mezzacappa_2014}. According to numerical simulations, most GW emission is expected to occur within approximately 1 second after the explosion, primarily due to conveciton and the standing accretion shock instability \citep{Blondin_apj2003, Andresen_mn2017, Muller2012AA, Kuroda2016ApJL, Morozova2018ApJ, Yakunin_2017, Vartanyan2023PhysRevD, Wang2023ApJ}.

In an ideal scenario, disregarding transient noise and glitches \citep{Nuttall2015CQGra}, the reconstruction of GW signals from CCSNe depends on our understanding of the signal's time-frequency structure \citep{Hayama_prd2015, Gossan2016PhRvD}. Since GW produced by CCSNe are highly stochastic and influenced by factors such as the progenitor star's mass, the equation of state of the resulting proto-neutron star \citep{Muller2016PASA, Morozova2018ApJ}. We cannot use matched filtering methods, as we do for searching GW from binary compact star mergers, to perform the search for GW from CCSNe \citep{Owen1999PhRvD, McIver2015}. For recent developments in the field, refer to the discussion in \cite{Andresen2024arXiv}. In recent years, various methods have been proposed and applied to detected reconstruct, and classify GW produced by CCSNe. These include wavelet analysis (which has been successfully used to reconstruct GW from binary compact star mergers) \citep{McIver2015, Mezzacappa2024arXiv}, ensemble empirical mode decomposition \citep{Yuan2024MNRAS, Hu2022ApJ}, principal component analysis \citep{Heng2009CQG, Rover2009PhRvd}, dynamic time warping \citep{Suvorova2019PhRvD}, and the increasingly popular machine learning techniques \citep{Chan2020PhRvD, Mitra2024MNRAS, Powell2024PhRvD}. However, these methods are primarily based on aLIGO and aVirgo and are used to reconstruct or classify GW waveforms from sources within 100 kpc.

In this paper, we explore the application of time-frequency analysis (TFA) to reconstruct the GW signal generated by CCSNe. The TFA method is an effective tool to characterize the time-varying features of non-stationary signals \citep{pons2014advanced, vedreno2013diagnosis}. The most classic TFA method is the short-time Fourier transform (STFT; \cite{gabor1946theory}). However, restricted by the Heisenberg uncertainty principle or unexpected cross-terms, the classical methods suffer from low time-frequency resolution, which leads to them not being able to characterize the non-linear behaviors of non-stationary signals precisely. Over the decades, TFA methods have continually evolved to address some of the limitations inherent in the STFT. Several advanced post-processing techniques have been proposed, including the reassignment method \citep{auger1995improving, auger2013time}, the synchrosqueezing transform (SST; \cite{daubechies2011synchrosqueezed, wang2013matching}), the multisynchrosqueezing transform (MSST; \cite{Yu2018MSST}), the improved multisynchrosqueezing transform (IMSST; \cite{Yu2020JSV}), parametric TFA method \citep{yang2011spline, yang2012multicomponent} and demodulated TFA \citep{wang2013matching, wang2013matchfault}. According to \cite{Yu2020JSV}, IMSST achieves high time-frequency resolution. In this study, we compare the advantages and disadvantages of two TFA methods, IMSST and STFT, in reconstructing GW signals generated by CCSNe, based on the capabilities of aLIGO and ET.

This paper is organized as follows. In Section \ref{sec:method}, we provide a brief overview of the basic concepts of TFA and its application in this study. In Section \ref{sec:sim}, we describe the method used to generate the simulated data. Section \ref{sec:rec} presents the results of reconstructing the simulated data using the IMSST method. In Section \ref{sec:fap}, we calculate the FAPR using the IMSST method to assess the likelihood that the reconstructed waveforms correspond to real signals. In Section \ref{sec:IMvsST}, we compare the performance of the IMSST and STFT methods in waveform reconstruction. Finally, a discussion of the results and future work is given in Section \ref{sec:dis}.

\section{Method}
\label{sec:method}

In this paper, we use two types of TFA methods (STFT and IMSST) to reconstruct the GW signals and compare the results from both methods. In this section, we briefly introduce these TFA methods.

\subsection{STFT}

The STFT is the most classical TFA method, which was proposed by \cite{gabor1946theory}. The STFT expression of a function with respect to the real and even window is defined as
\begin{equation}
G(\tau,\omega)=\int_{-\infty}^{+\infty}g(t-\tau)s(t)e^{-i\omega(t-\tau)}dt,
\label{eq:stft_f}
\end{equation}
where $g(t-\tau)$ is the moved window, $\omega$ is the angular frequency and $s(t)$ is the analyzed data. The STFT transforms a one-dimensional time-series signal into the two-dimensional time-frequency representation, allowing us to observe and extract the instantaneous amplitude and instantaneous frequency information of the signal. However, in both the time and frequency domain, the window function has a finite bandwidth, which results in an energy-blurred spectrogram $|G(\tau,\omega)|$. By integrating over the frequency direction, we derive the following expression:
\begin{equation}\begin{aligned}
&\int_{-\infty}^{+\infty}G(\tau,\omega)d\omega \\
&=\int_{-\infty}^{+\infty}\int_{-\infty}^{+\infty}g(t-\tau)\cdot s(t)\cdot e^{-i\omega(t-\tau)}dt d\omega \\
&=2\pi\cdot\int_{-\infty}^{+\infty}g(t-\tau)\cdot s(t)\cdot\delta(t-\tau)dt \\
&=2\pi g(0)\cdot s(\tau).
\end{aligned}\end{equation}

Therefore, the original signal $s(t)$ can be reconstructed by
\begin{equation}s(\tau)=\frac{1}{2\pi g(0)}\int_{-\infty}^{+\infty}G(\tau,\omega)d\omega.
\label{eq:stft_inv}
\end{equation}

\subsection{IMSST}

In this subsection, we briefly introduce the IMSST method, which is an improvement of the MSST method. More details can be found in \citep{Yu2018MSST, Yu2020JSV}. Because of restricted by the Heisenberg uncertainty principle or unexpected cross-terms, classical methods suffer from low time-frequency resolution. The IMSST, a high-resolution TFA tool based on multi-synchrosqueezing, addresses these issues and is effective for analyzing non-stationary signals.

In mathematics, the GW signal can be modeled as 
\begin{equation}
s(t) = A(t)e^{i\phi(t)},
\label{eq:st}
\end{equation}
where $A(t)$ represents the instantaneous amplitude, $\phi(t)$ denotes the instantaneous phase, and its first-order derivative $\phi^{'}(t)$, corresponds to the instantaneous frequency. The related STFT result for such a signal can be expressed as follows \citep{Yu2018MSST}:
\begin{equation}\begin{aligned}
&G(\tau,\omega)  \\
&\begin{aligned}&=A(\tau){e}^{{i}\varphi(\tau)}\int_{-\infty}^{+\infty}g(t-\tau){e}^{{i}(\varphi^{\prime}(\tau)(t-\tau))-{i}\omega(t-\tau)}\mathrm{d}u\\&=A(\tau){e}^{{i}\varphi(\tau)}{g}(\omega-\varphi^{\prime}(\tau)),\end{aligned} 
\end{aligned}
\label{eq:gt}
\end{equation}
where ${g}()$ denotes the Fourier transform of the window function. For any ($\tau$, $\omega$) where $|G(\tau,\omega)| \neq 0$, a 2-D instantaneous frequency estimate, $\hat{\omega}(\tau,\omega)$, can be obtained as
\begin{equation}\begin{aligned}
\hat{\omega}(\tau,\omega)& =\mathrm{Re}{\left(\frac{\partial_\tau G(\tau,\omega)}{{i}G(\tau,\omega)}\right)} \\
&=\mathrm{Re}\left(\frac{\partial_\tau(A(\tau)e^{i\varphi(\tau)}g(\omega-\varphi^{\prime}(\tau)))}{i(A(\tau)e^{i\varphi(\tau)}g(\omega-\varphi^{\prime}(\tau)))}\right) \\
& = \varphi^{\prime}(\tau),
\end{aligned}
\end{equation}
where Re() denotes the real part. To generate a concentrated time-frequency representation for time-varying signals, \cite{Yu2018MSST} proposed an iterative procedure to enhance the concentration of the time-frequency representation. The iterative procedure can be expressed as:
\begin{equation}
\begin{aligned}
&Ts^{[1]}(\tau,\eta) =\int_{-\infty}^{+\infty}G(\tau,\omega)\delta(\eta-\hat{\omega}(\tau,\omega))d\omega,  \\
&Ts^{[2]}(\tau,\eta) =\int_{-\infty}^{+\infty}Ts^{[1]}(\tau,\omega)\delta(\eta-\hat{\omega}(\tau,\omega))d\omega,  \\
&Ts^{[3]}(\tau,\eta) =\int_{-\infty}^{+\infty}Ts^{[2]}(\tau,\omega)\delta(\eta-\hat{\omega}(\tau,\omega))d\omega,  \\
&\text{:} \\
&Ts^{[N]}(\tau,\eta)=\int_{-\infty}^{+\infty}Ts^{[N-1]}(t,\omega)\delta(\eta-\hat{\omega}(\tau,\omega))d\omega .
\end{aligned}
\end{equation}
This can be expressed as
\begin{equation}
Ts^{[N]}(\tau,\eta)=\int_{-\infty}^{+\infty}G(\tau,\omega)\delta(\eta-\hat{\omega}^{[N]}(\tau,\omega))d\omega,
\end{equation}
where $\hat{\omega}^{[2]}(\tau,\omega)=\hat{\omega}(\tau,\hat{\omega}(\tau,\omega))$, $\hat{\omega}^{[3]}(\tau,\omega)=\hat{\omega}(\tau,\hat{\omega}(\tau,\hat{\omega}(\tau,\omega)))$, and so on.

However, in the IMSST method, to address the issue of non-reassignment, the GW signal can be modeled as:
\begin{equation}
s(t) = A(t)e^{i(\phi(t) + \phi^{'}(t)(u-t)+0.5\phi^{''}(u-t)^2)},
\label{eq:imsstst}
\end{equation}
and the window function is a Gaussian function:
\begin{equation}
g(t) = e^{-\frac{t^2}{2\sigma}},
\end{equation}
where $\sigma$ is the standard deviation of the Gaussian function. Therefore, 
\begin{equation}
\begin{aligned}
    G(\tau,\omega) &= A(t)e^{i\varphi(t)} \int_{-\infty}^{+\infty}  e^{-(2\sigma)^{-1}(1-i\sigma\varphi^{\prime\prime}(\tau))(t-\tau)^2} \\
    &\hspace{2cm} \times e^{-i(\omega-\varphi^{\prime}(\tau))(t-\tau)} \, d(t-\tau) \\
    &= A(\tau)e^{i\varphi(\tau)} \sqrt{\frac{2\sigma\pi}{1-i\sigma\varphi^{\prime\prime}(\tau)}} e^{-\frac{\sigma(\omega-\varphi^{\prime}(\tau))^2}{2-i2\sigma\varphi^{\prime\prime}(\tau)}}.
\end{aligned}
\end{equation}

In the IMSST method, the instantaneous frequency estimate $\omega^{[N]}(\tau, \omega)$ is obtained through two rounding operations. At $\tau_0$, $G(\tau_0, \omega)$ is reassigned to a single frequency point, which effectively addresses the blurry energy problem of the MSST. The related expression is given by: 
\begin{equation}
Ts^{[N]}(\tau,\eta)=\int_{-\infty}^{+\infty}G(\tau,\omega)\delta(\eta-\hat{\omega}_R^{[N]}(\tau,\omega))d\omega,
\end{equation}
where $\hat{\omega}_R^{[N]}(\tau,\omega)$ represents the frequency after two rounds of operation, given by ($\hat{\omega}_R^{[N]}(\tau,\omega)=\text{Round}(\text{Round}(2\hat{\omega}^{[N]}(\tau,\omega))/2)$). The following expressions demonstrate that the original signal can be perfectly reconstructed from the IMSST results:
\begin{equation}\begin{aligned}
&\int_{-\infty}^{+\infty}Ts^{[N]}(\tau,\eta)d\eta \\
&=\int_{-\infty}^{+\infty}\int_{-\infty}^{+\infty}G(\tau,\omega)\delta(\eta-\hat{\omega}_R^{[N]}(\tau,\omega))d\omega\mathrm{d}\eta \\
&=\int_{-\infty}^{+\infty}G(\tau,\omega)\int_{-\infty}^{+\infty}\delta(\eta-\hat{\omega}_R^{[N]}(\tau,\omega))\mathrm{d}\eta d\omega \\
&=(2\pi g(0))s(\tau).
\end{aligned}\end{equation}

Thus, the reconstructed GW signal can be expressed as:
\begin{equation}
s(\tau)=(2\pi g(0))^{-1}\int_{-\infty}^{+\infty}Ts^{[N]}(\tau,\omega)d\omega .
\end{equation}

\section{simulations and results}

\subsection{Data simulations}
\label{sec:sim}

In this section, we describe the testing phase designed to evaluate the effectiveness of the IMSST and STFT TFA methods in enhancing waveform reconstruction, as well as their performance in aLIGO and ET. As a preliminary exploration, this paper focuses exclusively on a single aLIGO and ET detector to assess the feasibility of these two TFA approaches.

As a preliminary step, we utilize simulated waveforms derived from various sources in the literature. The magnetorotational CCSN signals are sourced from \citep{Abdikamalov2014PhRvD, Dimmelmeier2008PhRvD, Richers2017PhRvD}\footnote{\url{https://stellarcollapse.org/gwcatalog.html}, \url{https://zenodo.org/record/201145.}}, with the GW waveforms generated by \cite{ Dimmelmeier2008PhRvD} based on 2-D simulations. For the neutrino-driven mechanism, we employ the waveforms from \citep{Yakunin_2017,  Ott2013ApJ, Andresen2019MNRAS, Kuroda2017ApJ, Muller2012AA, Powell2019MNRAS, Radice2019ApJL, Powell2020MNRAS, Vartanyan2023PhysRevD}\footnote{\url{https://wwwmpa.mpa-garching.mpg.de/ccsnarchive/data/Andresen2019/}, \url{https://www.astro.princeton.edu/~burrows/gw.3d/}}. These simulations cover a wide range of progenitor masses, from 9 $M_\odot$ to 60 $M_\odot$, as shown in Table \ref{tab:waveforms}. Since these waveforms are generated under varying conditions, such as differing distances, sampling rates, and durations, normalization is necessary before generating the time series. To achieve this, we scale the amplitudes of the waveforms by relocating the sources to a distance of 10 kpc from Earth. Additionally, to ensure uniformity in the sampling rates, all waveforms are down-sampled to a preselected rate of 4096 Hz. It is important to noted that some of the waveforms used in this study are generated under axisymmetric simulations, in which the waveforms are described solely by the polarization $h_+(t)$, with the corresponding $h_\times(t)$ components defined as vectors of zeros.

The subsequent step involves simulating the data observed by aLIGO and ET as described in Eq (\ref{eq:data}),
\begin{equation}
d(t)=F_{+}(\alpha, \delta)h_{+}(t)+F_{\times}(\alpha, \delta)h_{\times}(t)+n(t) = h(t) + n(t),
\label{eq:data}
\end{equation}
where $n(t)$ represents the instrument noise. For the aLIGO simulation, we utilize the zero detuning, high power noise power spectral density (PSD) \citep{Abbott2020LRR}. In contrast, for ET, the noise PSD function corresponding to the ET-D configuration is adopted \citep{Hild2011CQGra}. The strain signals $h_{+}(t)$ and $h_{\times}(t)$ correspond to the two independent polarization modes, while $F_{+}(\alpha, \delta)$ and $F_{\times}(\alpha, \delta)$ denote the antenna pattern functions ($\alpha$ is right ascension and $\delta$ is declination). A random location of ($\alpha$, $\delta$) in the sky is selected from a uniform distribution on ($\alpha$, $\sin \delta$). Given the short duration of the simulated signals (approximately 1 second), it is reasonable to assume that both antenna pattern functions remain time-independent and therefore be treated as real constants. Although the spatial positions of these sources are randomly distributed in the sky, their relative positions with respect to aLIGO and ET remain constant.

Before reconstructing the GW signals using TFA methods, preprocessing is a essential step. This involves performing a spectral analysis, whitening the detector's colored noise to approximate Gaussian white noise, and applying bandpass filter with a frequency range of [20, 2000] Hz. The resulting preprocessed data, after whitening and bandpass filtering, is denoted as $\tilde{d}(t)$. These steps effectively ensure that the noise is transformed to more closely resemble Gaussian white noise, facilitating more robust signal reconstruction. For further analysis, we extract simulated data within a time window of approximately 1 second. 

\begin{table*}
\centering
\caption{The mass ranges and mechanisms of the progenitors associated with the simulated waveforms used in this work. The first column lists the corresponding study. The "Mechanism" column indicates the explosion mechanism for each waveform, with "M" representing the magnetorotational mechanism and "N" indicating the neutrino-driven mechanism. The labels M1, M2, N1, and N2 refer to the specific waveforms selected for our case studies. The "Mass" column specifies the progenitor masses in solar mass units as represented in the simulations. The "No." column denotes the number of waveforms available from each study. All listed masses correspond to the stellar masses at zero age unless otherwise noted.}\label{tab:data}
\begin{tabular}{llll}
\hline
\hline
             & Mechanism & Mass ($M_{\odot}$)                       & No.  \\ \hline
Abdikamalov \citep{Abdikamalov2014PhRvD}  & M (M1)        & 12.0                          & 92   \\ 
Dimmerlmeier \citep{Dimmelmeier2008PhRvD} & M        & 11.2, 15.0, 20.0, 40.0        & 136  \\ 
Richers \citep{Richers2017PhRvD}      & M (M2)       & 12.0                          & 1824 \\ 
Andresen \citep{Andresen2019MNRAS}     & N        & 11.2, 15.0                    & 6    \\ 
Kuroda \citep{Kuroda2017ApJ}       & N        & 11.2, 15.0                    & 2    \\ 
Muller \citep{Muller2012AA}       & N        & 15.0, 20.0                    & 6    \\ 
Ott \citep{Ott2013ApJ}          & N        & 27                            & 8    \\ 
Powell \citep{Powell2019MNRAS}       & N        & 18.0                     & 1    \\ 
Powell \citep{Powell2020MNRAS}       & N (N1)       &  39.0                     & 1    \\ 
Radice \citep{Radice2019ApJL}       & N        & 9, 10, 11, 12, 13, 19, 25, 60 & 8    \\ 
Yakunin \citep{Yakunin2017arXiv}      & N (N2)       & 15                            & 1    \\ 
Vartanyan \citep{Vartanyan2023PhysRevD}  &N    & 9, 9.25, 9.5, 11, 12.25, 14, 15.01, 23 & 11   \\
\hline
\hline\\

\end{tabular}

\label{tab:waveforms}
\end{table*}

\subsection{Reconstruction}
\label{sec:rec}

\begin{figure*}

     \includegraphics[width=7.2cm]{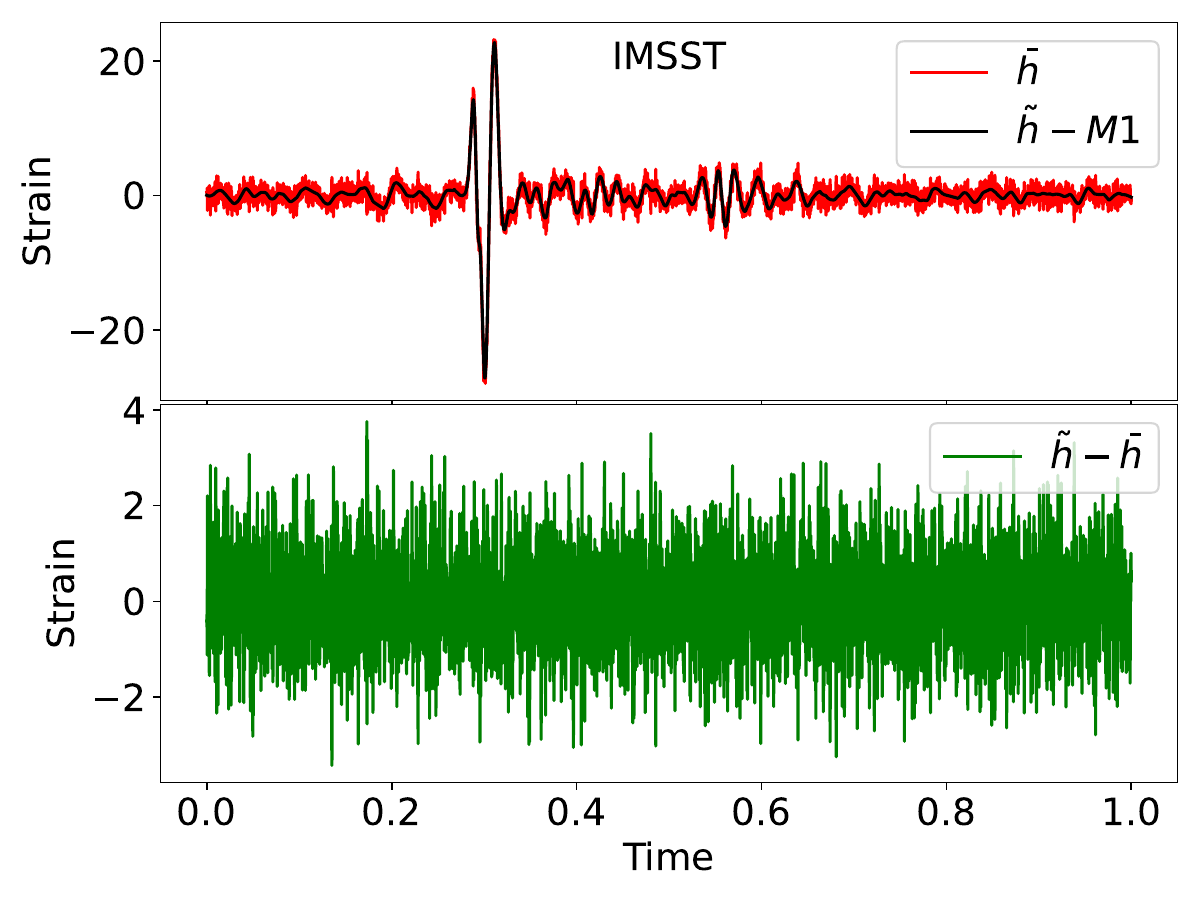}
     \includegraphics[width=7.2cm]{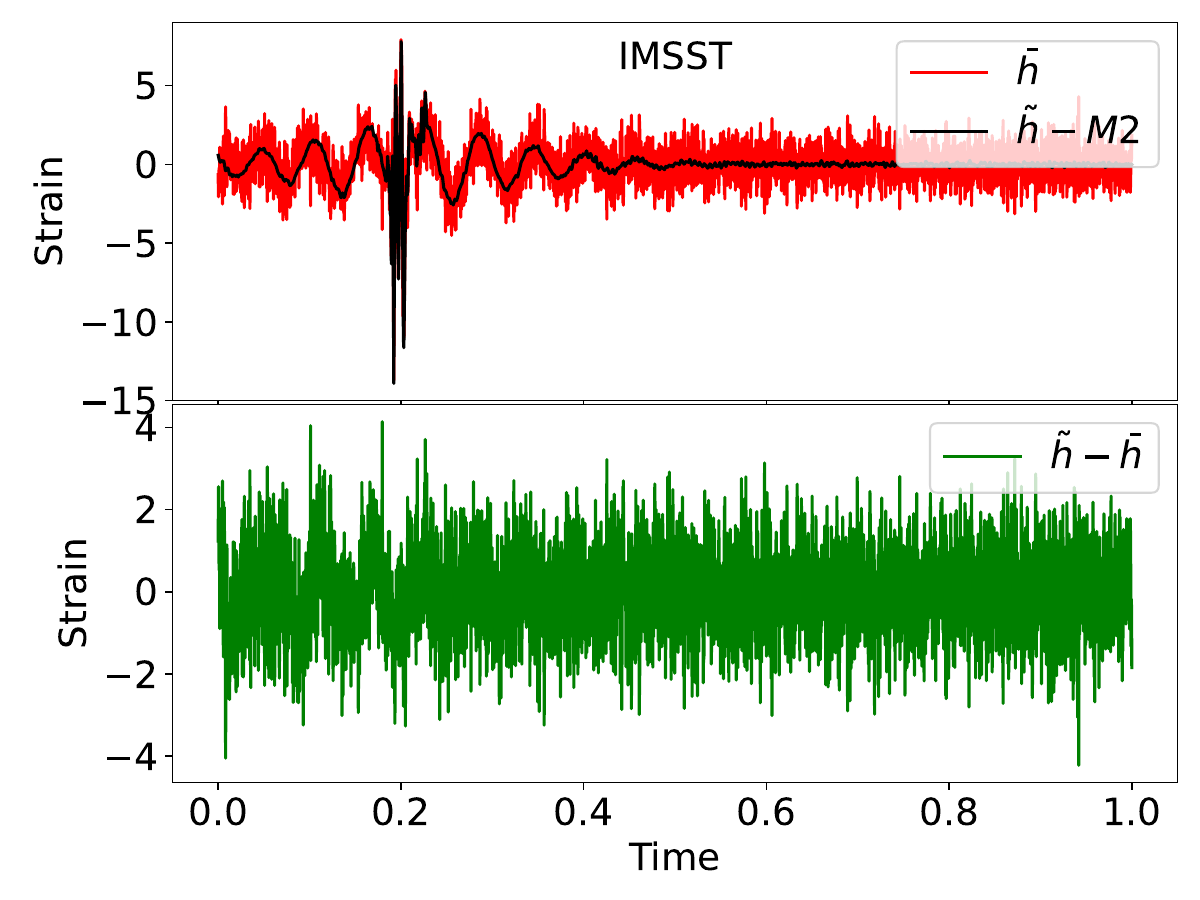}

    \includegraphics[width=7.2cm]{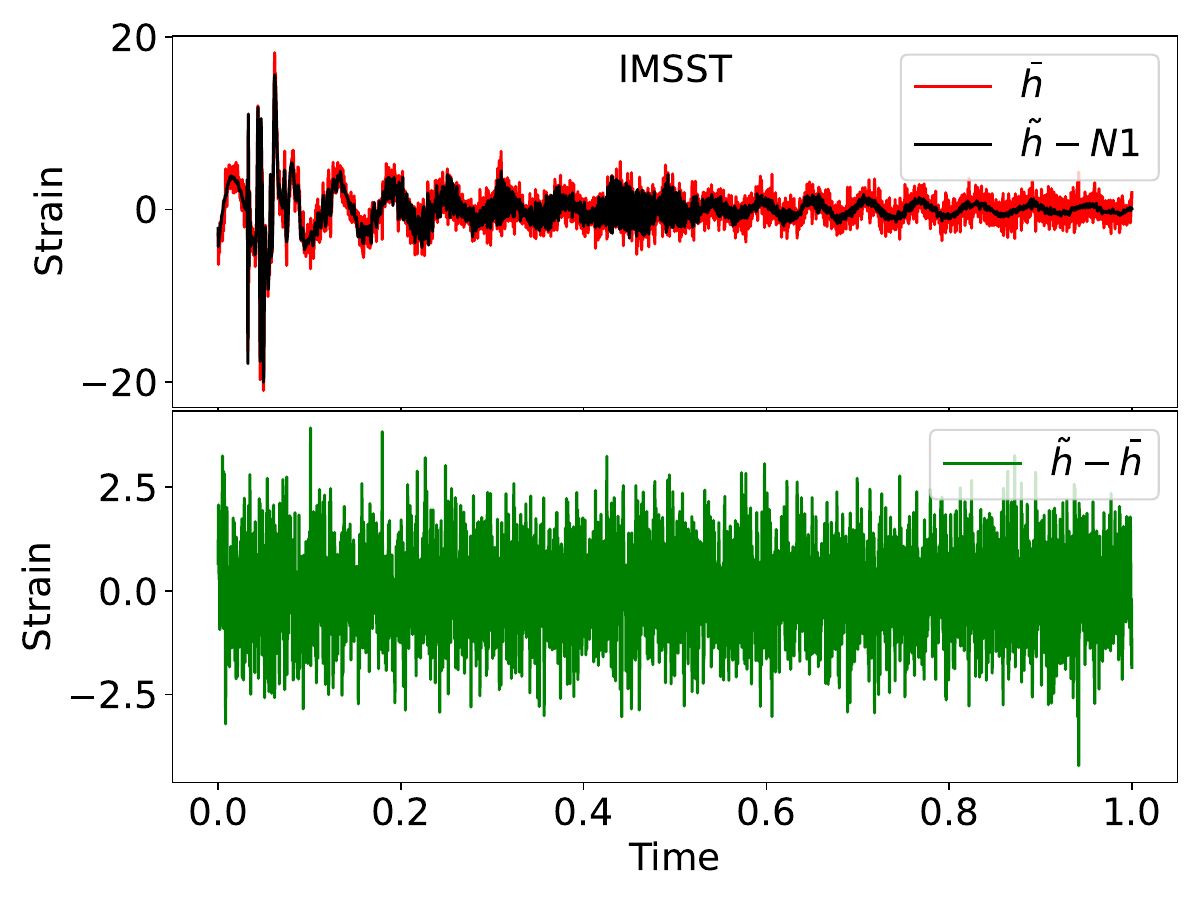}
   \includegraphics[width=7.2cm]{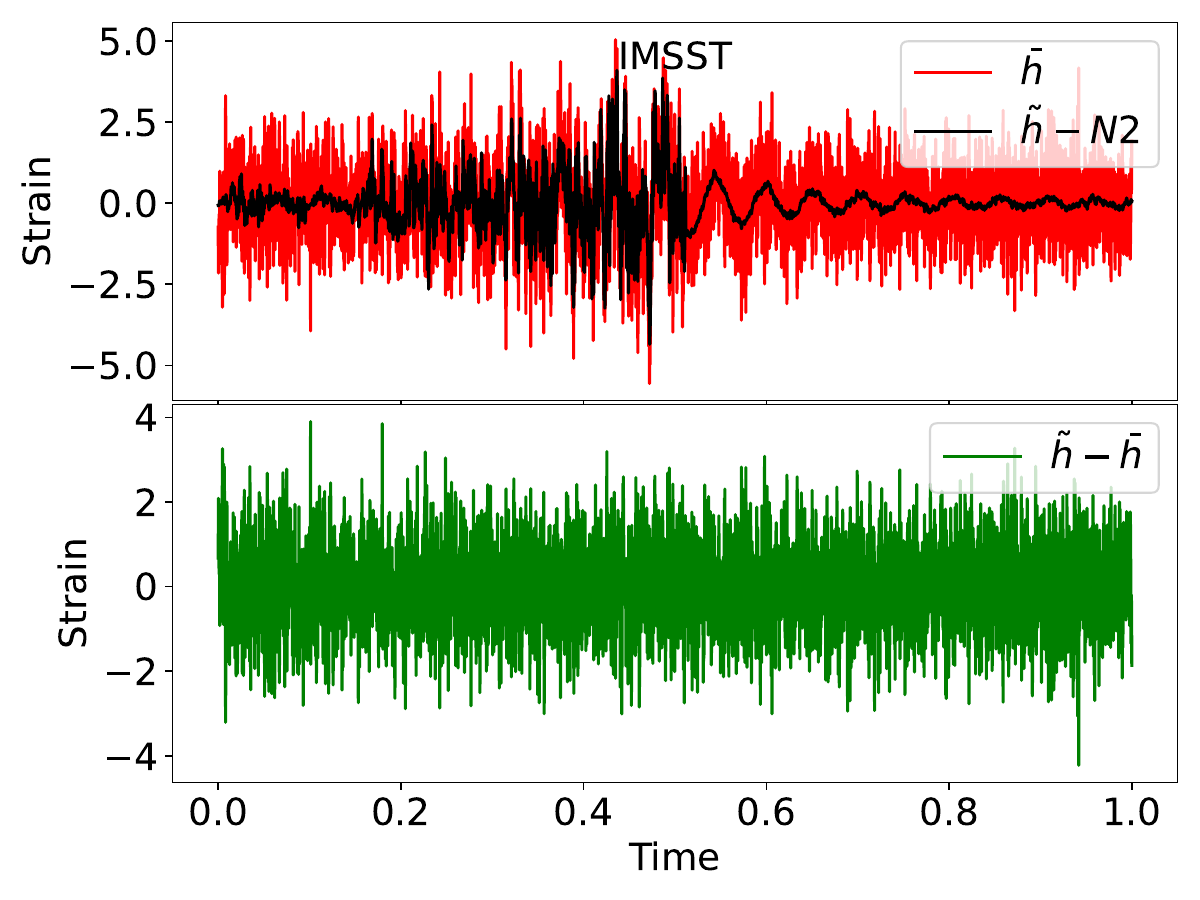} 

\caption[]{Results of the waveform reconstruction at a distance of 10 kpc using the IMSST method for four GW signals. In each panel, the top graph represents the whitened and filtered injected waveform (black line) and the reconstructed waveform (red line), the bottom graph represents the residual between the whitened and filtered injected waveform and the reconstructed waveform (green line).
 \label{fig:rec_h}}
\end{figure*}

We have selected four waveforms from Table \ref{tab:waveforms} that exhibit significant difference in both time-domain and frequency-domain structures as case studies for illustrating these TFA methods.
The two waveforms, labeled as M1 and M2, are taken from \cite{Abdikamalov2014PhRvD, Richers2017PhRvD} and are generated by the magnetorotational mechanism. These two waveforms have relatively simple time-domain structures, containing of several prominent peaks, with a frequency distribution in the range of 20-200 Hz (as indicated by the black lines in the upper panels of Fig. \ref{fig:rec_h}). The other two waveforms, labeled as N1 and N2, are taken from \cite{Powell2020MNRAS, Yakunin2017arXiv} and are generated by neutrino-driven mechanism (as shown by the black lines in the lower panels of Fig. \ref{fig:rec_h}). Compared to M1 and M2, N1 and N2 exhibit more complex time-domain structures, feature not only more peaks but also rapid oscillations, resulting in higher frequency components. The frequency distribution range is 20-1200 Hz. Using these four simulated GW data, we present the results of reconstructing GW signals using the IMSST methods, as shown in Fig. \ref{fig:rec_h}. From the results in Fig. \ref{fig:rec_h}, it can be observed that 
the IMSST method effectively reconstructs the main features of the injected waveforms, even though the reconstructed waveforms still contain some noise components. Additionally, it can be concluded that when the amplitude of the injected waveform is larger, the proportion of noise components in the reconstructed waveform becomes smaller, as shown in the results on the left side of Fig. \ref{fig:rec_h}.

We aim to assess the accuracy of the reconstruction by evaluating the degree of concordance between the reconstructed GW waveform and the injected GW waveform. We will continue to use the match score ($\eta$) as a metric to measure the accuracy of the reconstruction \citep{Suvorova2019PhRvD, Yuan2024MNRAS}:
\begin{equation}
\eta = \frac{(\tilde{h}|\bar{h})}{\sqrt{(\tilde{h}|\tilde{h})(\bar{h}|\bar{h})}},
\label{match_score}
\end{equation}
where $\tilde{h}$ is injected waveform after bandpass filtering, $\bar{h}$ is the reconstructed waveform by the TFA method, and $(a|b)$ denoted the inner product\footnote{
In \cite{Suvorova2019PhRvD}, the inner product includes a noise weighting term, equivalent to whitening the data. In our processing, we have already whitened the data, so there is no need to apply noise weighting to the inner product here.}.

\begin{figure}
\centering
    \includegraphics[width=8.5cm]{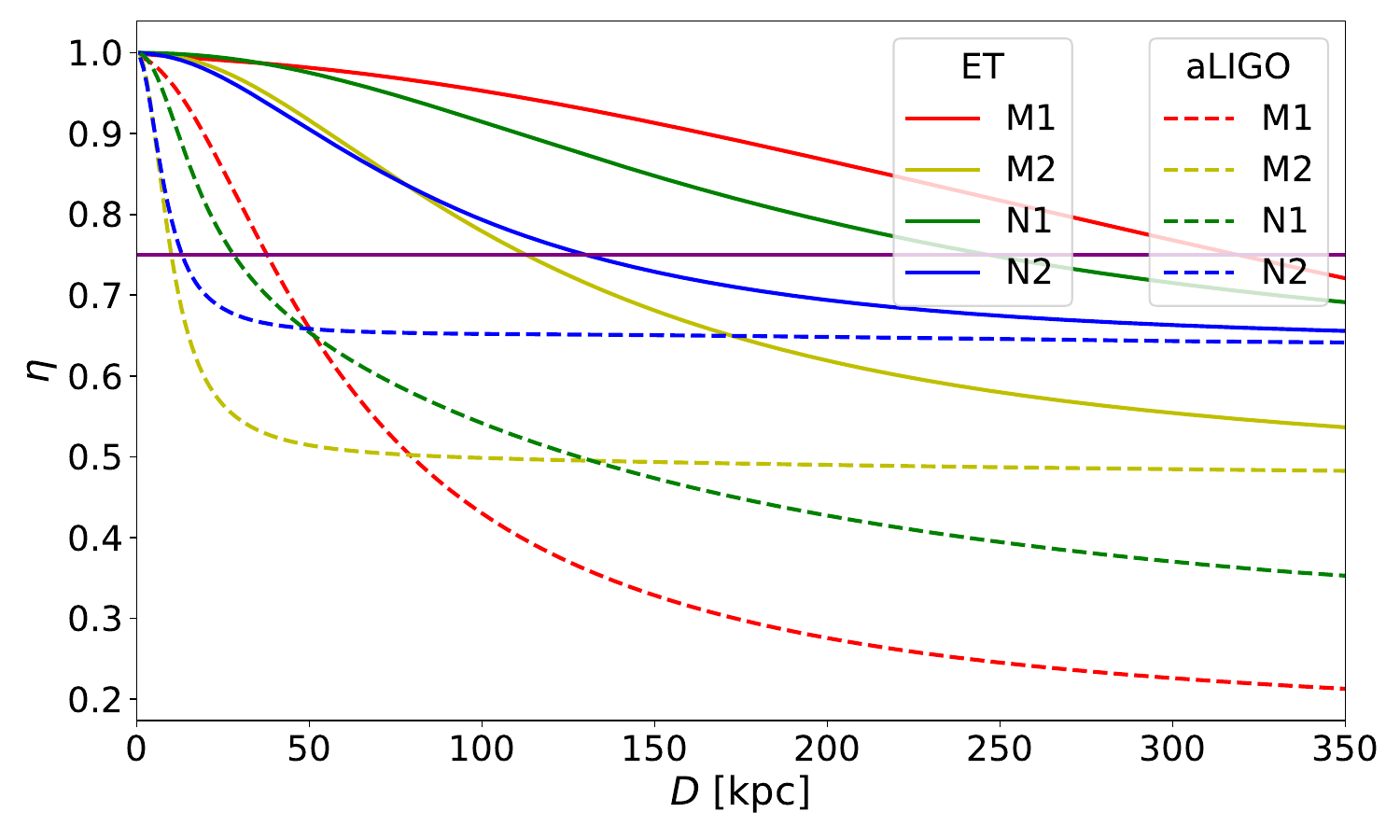}
\caption[]{The evolution of the 
match score with distance. The IMSST analysis results based on the simulated data from ET and aLIGO detectors are represented by solid and dashed lines, respectively. The four colors correspond to four different GW waveforms: red for M1, yellow for M2, green for N1, and blue for N2. The purple
horizontal line represents the threshold, which is 0.75.}
\label{fig:distance}
\end{figure}

In the subsequent steps, we evaluate the capabilities of the IMSST method in reconstructing GW waveforms for the aLIGO and ET detectors, with a focus on its performance at various distances. To achieve this, we simulated GW observation data for aLIGO and ET at different distances,  keeping the spatial location of the GW source constant across detectors, and applied the IMSST methods for waveform reconstruction. Additionally, we calculate the match score between the reconstructed waveforms and the injected waveforms, using the four waveforms illustrated in Fig. \ref{fig:rec_h} as examples, with results presented in Fig. \ref{fig:distance}. The results in Fig. \ref{fig:distance} reveal that, regardless of the specific waveform, the match score decrease as the distance increases. This decline is attributed to the reduce strain of GW signals at greater distance, which increasing the noise in waveforms reconstructed by the IMSST methods. However, the rate of decrease in the match score eventually stabilizes. This trend is primarily due to the partial alignment between the extracted data and the injected waveforms, suggesting that variation in the match score is not solely dependent on distance.

To determine whether the GW waveform has been successfully reconstructed, it is necessary to establish at a threshold match score. In this study, we adopt the threshold of 0.75, which was set in previous research \citep{Yuan2024MNRAS}. A match score greater than or equal to 0.75 indicates that the GW waveform has been successfully extracted and reconstructed from the observed data. Conversely, a score below 0.75 suggests that the reconstructed waveform primarily consists of noise. The results presented in Fig. \ref{fig:distance} show a comparison of the IMSST method's reconstruction capability for the ET and aLIGO detectors. Assuming the reconstructed signal is indeed a GW, IMSST can reconstruct the GW waveform at maximum distance of 317 kpc and 37 kpc for the ET and aLIGO detectors, respectively. This result is consistent with expectations, as the sensitivity of the ET detector is approximately one order of magnitude higher than that of aLIGO, allowing for the reconstruction of GW signals at greater distances. Additionally, the higher sensitivity of the ET detector means it has a lower noise level, which causes the match score to decline more slowly with increasing distance compared to aLIGO. We also observed that, regardless of whether the ET or aLIGO detector was used, when the match score exceeds the set threshold, the match score for M1 and N1 are significantly higher than those for M2 and N2. This difference can be attributed to the significantly higher strain of M1 and N1 at the same distance, enabling the IMSST method to more effectively reconstruct the GW waveforms from the data. This finding is further corroborated by the results shown in Fig. \ref{fig:rec_h}.


\subsection{False alarm rate of reconstruction}
\label{sec:fap}

Additionally, we calculate the match score between the reconstructed waveform, derived from without any injected GW waveforms, and the injected GW waveforms. This analysis is conducted to validate the accuracy of the reconstructed waveform and to mitigate the risk of misclassifying noise as a signal. We determine the threshold $\eta$ from the cumulative distribution of the match score for the noise-only realization $P_{\text{noise}}$, enabling the calculation of the false alarm probability of reconstruction (FAPR) as presented in Eq (\ref{eq:FAP}):
\begin{equation}
1-P_{\text{noise}}(\eta) \leq \text{FAPR},
\label{eq:FAP}
\end{equation}

\begin{figure}
\centering
    \includegraphics[width=8.5cm]{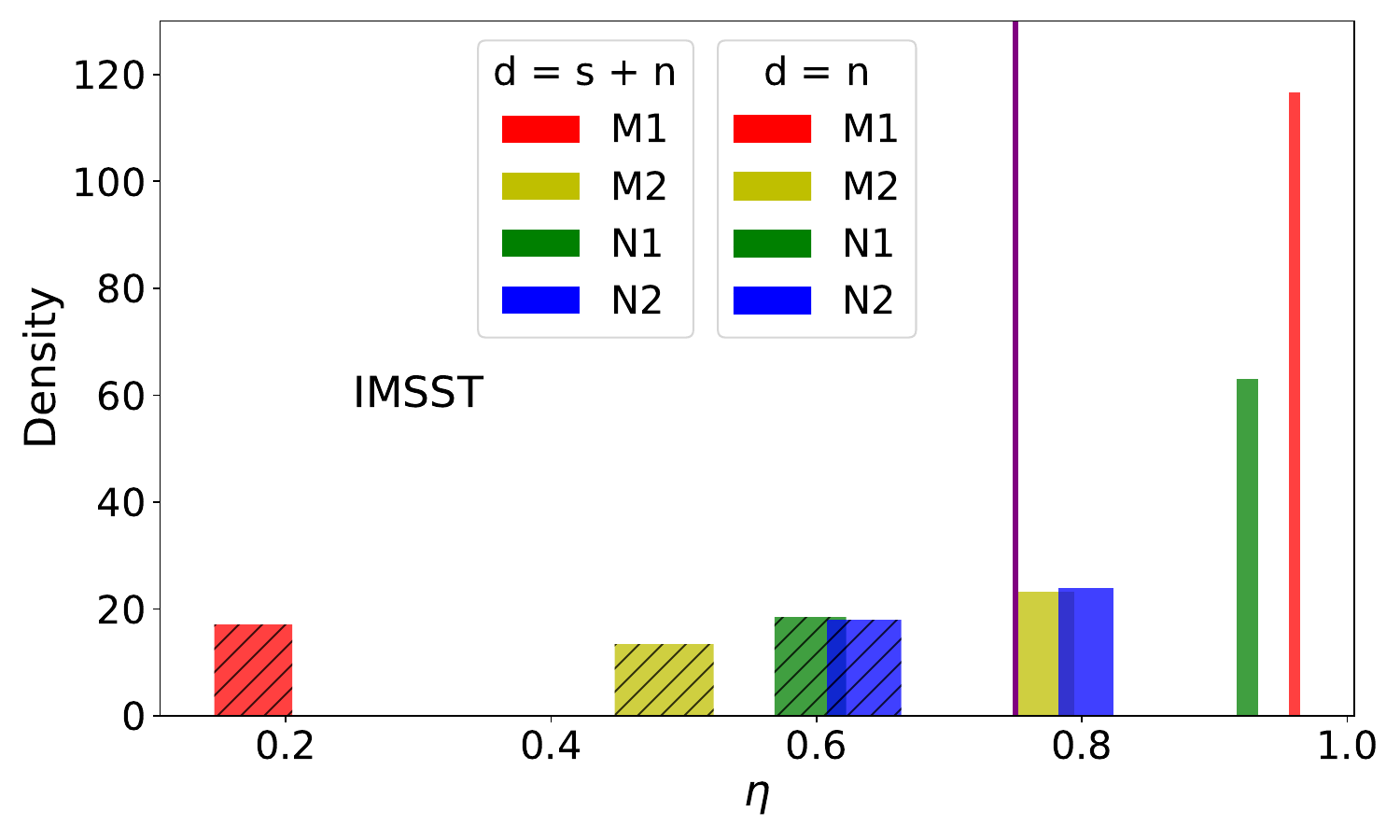}
\caption{Histogram of match scores for the simulated data based on aLIGO detector at 10 kpc, illustrating the distributions with waveform injection (solid-colored bars) and without waveform injection (hatched bars). The injected waveforms include M1, M2, N1, and N2, each occurring with equal frequency. The purple vertical line indicates the threshold of 0.75.}
\label{fig:stability}
\end{figure}

Based on the aLIGO detector, at 10 kpc, we generated 100 simulated data sets for each of the aforementioned four waveforms, resulting in a total of 400 simulated data sets with injected waveforms, and calculated their match score distributions. Additionally, we generated 400 sets of pure noise data, applied the IMSST methods for reconstruction, and averaged them into four groups. Each group was then matched with the four waveforms to obtain the distribution of match scores without waveform injection. The results are shown in Fig. \ref{fig:stability}.
When the threshold $\eta = 0.75$, the FAPR for the IMSST method at 10 kpc is less than $2.5 \times 10^{-3}$. This shows that the TFA method can effectively distinguish between GW signals and pure noise. It confirms the method's reliability and effectiveness in reconstructing GW signals. Furthermore, the results shown in Fig. \ref{fig:stability} reveal that, in the distribution of match score without waveform injection (hatched bars), approximately one-quarter of the results are distributed around 0.2. This is primarily due to the lower match score between the reconstructed waveform without injected waveforms and the M1 waveform. Compared to the other three waveforms, the M1 waveform has simpler components and a larger amplitude, which results in a lower match with noise and a lower match score. In the match score distribution for the injected waveforms (solid-colored bars), low match scores are mainly due to the M2 and N2 waveforms, which have smaller amplitudes. This result is also consistent with the findings in Fig. \ref{fig:rec_h} and Fig. \ref{fig:distance}. Therefore, the match score for these waveforms are lower than for the others. Based on these results, we conclude that the TFA methods for reconstructing GW waveforms are related to their amplitudes, but not to the mechanisms of waveform generation. Thus, the current match scores are not suitable for classifying GW waveforms directly.


\begin{figure}
\centering
    \includegraphics[width=8.5cm]{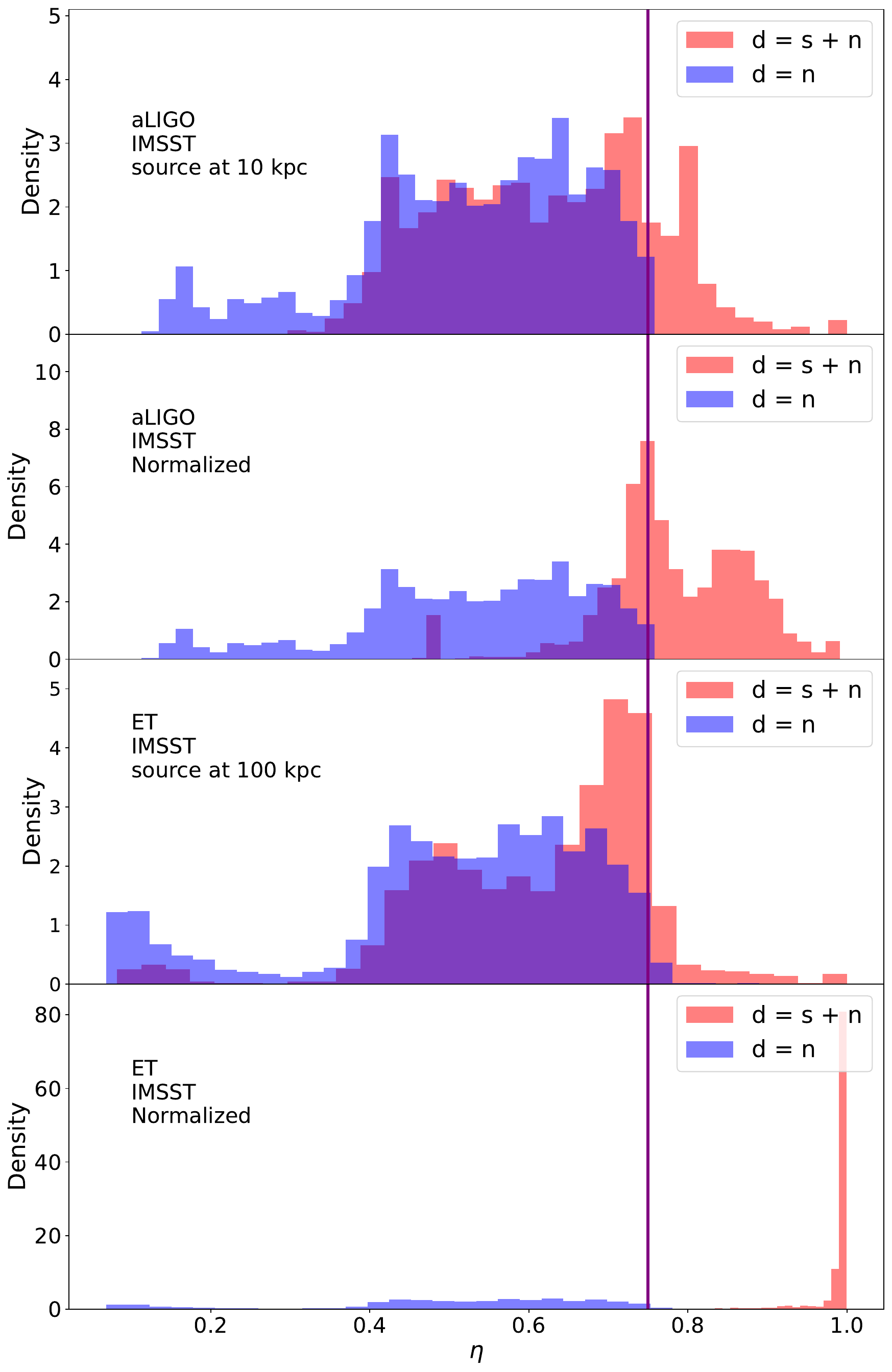}
\caption{The histograms of the match scores for the simulated datasets based on the aLIGO and ET detectors, including the distributions for match scores with waveform injection (blue) and without waveform injection (purple). The injected waveforms consist of the 2096 waveforms listed in Table \ref{tab:data}, with each waveform injected an equal number of times. The purple vertical line indicates the threshold of 0.75. The first and second panels show the match score distribution results for the simulated data at a distance of 10 kpc using the IMSST method, with the first panel corresponding to the unnormalized data and the second panel corresponding to the data with waveform amplitudes normalized to $5 \times 10^{-21}$. The corresponding FAR values are $2.1 \times 10^{-2}$ and $6.2 \times 10^{-3}$, respectively. The third and fourth panels show the match score distribution results for the simulated data based on the ET detector, with the third panel corresponding to the data at a distance of 100 kpc, and the fourth panel corresponding to the data with waveform amplitudes normalized to $5 \times 10^{-21}$. The corresponding FAR values are also $1.3 \times 10^{-1}$ and $1.5 \times 10^{-2}$, respectively.}
\label{fig:RFAP_LIGO}
\end{figure}

Finally, we analyzed the 2096 waveforms from Table \ref{tab:data}. Each waveform was injected twice at a distance of 10 kpc, resulting in a total of 4170 simulated datasets, which were generated based on the aLIGO detector's simulation. Additionally, we generated 4170 pure noise datasets (i.e., without waveform injection) and processed both the injected and non-injected datasets using the IMSST method. The corresponding match score obtained from these analyses are shown in the first panel of Fig. \ref{fig:RFAP_LIGO}. With a threshold set to $\eta = 0.75$, the FAPR value at 10 kpc for IMSST was $2.1 \times 10^{-2}$. We also performed calculations similar to those shown in the top panel of Fig. \ref{fig:RFAP_LIGO}, but this time normalizing the maximum amplitude to $5 \times 10^{-21}$. Using the IMSST method, we repeated the same procedure on the simulated data, both with and without waveform injection, yielding the FAPR distributions shown in the second panel of Fig. \ref{fig:RFAP_LIGO}. Under the normalized conditions, with a threshold of $\eta = 0.75$, the FAPR value obtained from the IMSST analysis was $6.2 \times 10^{-3}$. For the datasets generated based on the ET detector, the total number of datasets is the same as for the aLIGO detector, with equal numbers of datasets containing and not containing waveform injections. The spatial location of the GW source is kept constant across both detectors. Since the sensitivity of the ET detector is approximately one order of magnitude higher than that of aLIGO, we placed the GW source at a distance of 100 kpc and analyzed these datasets using the IMSST method. With the same threshold, the resulting FAPR value was $1.3\times10^{-1}$, as shown in the third panel of Fig. \ref{fig:RFAP_LIGO}. To eliminate the additional effects of amplitude, we normalized the templates by scaling the maximum amplitude of the waveforms to $5 \times 10^{-21}$. Under this normalization, we performed the same analysis on the newly generated datasets using the IMSST method. The resulting FAPR value, with the same threshold, was $1.5 \times 10^{-2}$, as shown in the fourth panel of Fig. \ref{fig:RFAP_LIGO}. From the results shown in Fig. \ref{fig:RFAP_LIGO}, it can be seen that for both the aLIGO and ET detectors, the reconstruction event rate after waveform normalization is higher than the rate without normalization. This is because some GW waveforms have amplitudes below $\text{max}(h(t)) < 5 \times 10^{-21}$, and normalization effectively places these sources at closer distances, thereby increasing the reconstruction event rate.
However, normalization does not ensure successful reconstruction of all waveforms. Some GW sources are outside the optimal sky regions for detector response, leading to lower detected amplitudes that cannot be reconstructed. Additionally, from Fig. \ref{fig:RFAP_LIGO}, it is evident that compared to the match score distributions for non-injected waveforms in both aLIGO and ET, the minimum match score for ET is around 0.1, while for aLIGO it is around 0.2. This difference is closely related to the sensitivity of the detectors. Since ET has a lower noise level, the match score obtained without signal injection are correspondingly lower.

\subsection{Comparison of IMSST and STFT methods}
\label{sec:IMvsST}

In this subsection, we compare the performance of the IMSST method and the STFT method under the ET detector, focusing on the differences in the maximum detection distance and the FAPR.

\begin{figure}
\centering
    \includegraphics[width=8.5cm]{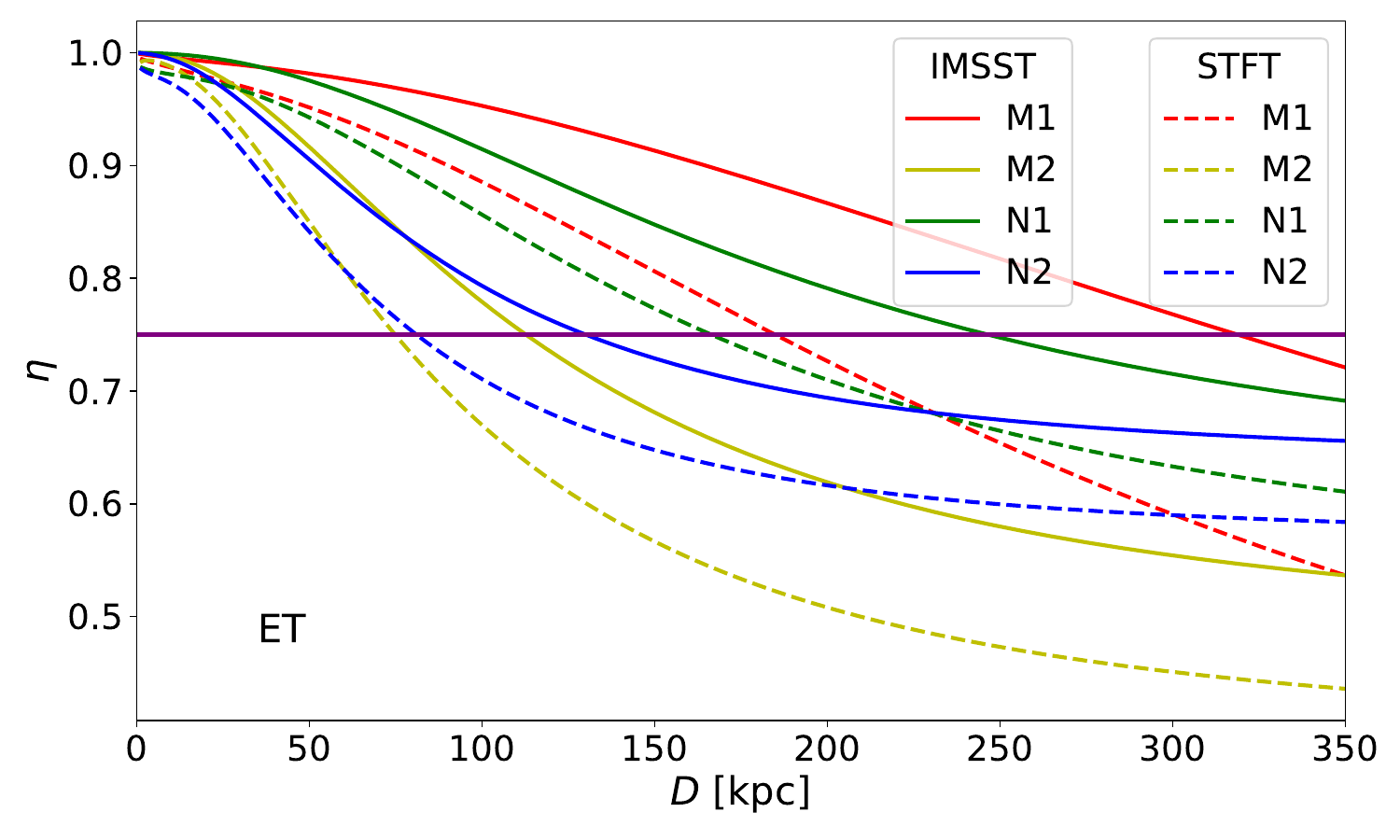}
\caption[]{The evolution of the match score with distance for the two time-frequency analysis methods. The IMSST and STFT analysis results based on the simulated data from the ET detector are represented by solid and dashed lines, respectively. The four colors correspond to four different gravitational wave waveforms: red for M1, yellow for M2, green for N1, and blue for N2. The purple horizontal line represents the threshold, which is 0.75.}
\label{fig:IM_ST_distance}
\end{figure}

We continue using the four waveforms from Fig. \ref{fig:distance}, based on simulated data generated with the ET detector, to compare the differences in reconstructible distance between the IMSST method and the STFT method, while maintaining a reconstruction threshold of 0.75. The evolution of the match score with distance for the two TFA methods is shown in Fig. \ref{fig:IM_ST_distance}. Assuming all reconstructed waveforms are real GW signals, it can be observed from Fig. \ref{fig:IM_ST_distance} that the maximum reconstructable distance for the IMSST method (317 kpc) is greater than that for the STFT method (186 kpc). As the distance increases, the match score decreases for both methods, but the decrease is more pronounced for the STFT method, with the match score eventually leveling off. This decreasing trend is consistent with the results shown in Fig. \ref{fig:distance}.

\begin{figure}
\centering
    \includegraphics[width=8.5cm]{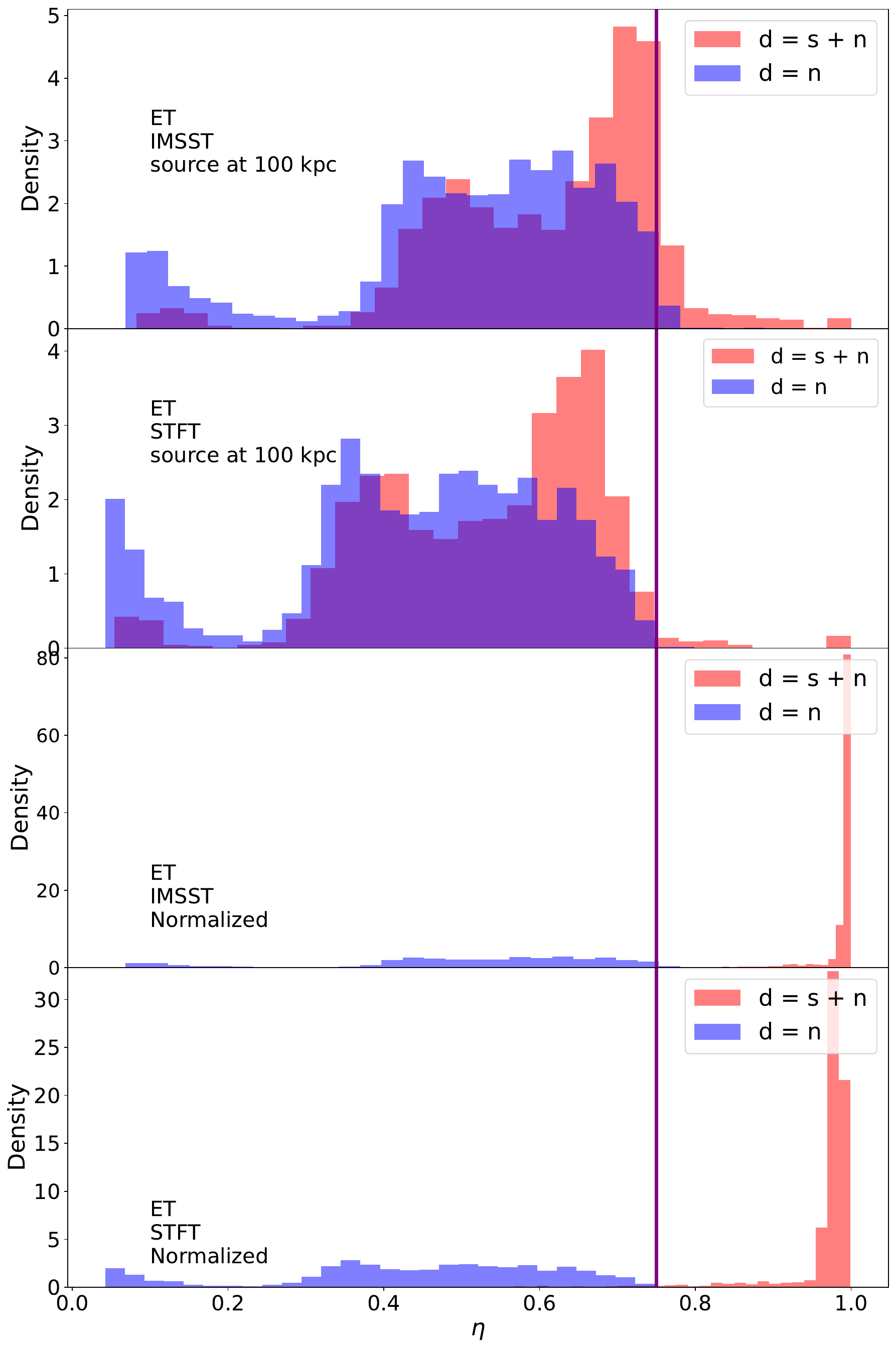}
\caption{The panels from the first to the fourth row are similar to those in Fig. \ref{fig:RFAP_LIGO}, but the simulated data are based on ET detector. The first and second rows of the graph show the match score distribution results obtained from analyzing the simulated data at 100 kpc using the IMSST and STFT methods, with corresponding FAR values of $1.3\times10^{-1}$ and $5.6\times10^{-2}$, respectively. The third and fourth rows of the graph show the match score distribution results for simulated data with strain normalized to $5\times10^{-21}$ using the IMSST and STFT methods, with corresponding FAR values of $1.5\times10^{-2}$ and $9.7\times10^{-4}$, respectively.}
\label{fig:RFAP_ET}
\end{figure}

Next step, we compare the differences in FAPR between the two TFA methods. To evaluate the performance of these methods with the ET detector, we used a simulated dataset based on the ET detector. This dataset consists of 4170 simulated datasets and 4170 pure noise datasets. We applied both IMSST and STFT methods to analyze these datasets and computed the corresponding match score. The results are shown in Fig. \ref{fig:RFAP_ET}. In the simulation, we first set the source location to 100 kpc. With a threshold of $\eta = 0.75$, the FAPR values for the IMSST and STFT methods on the ET simulated data were $2.1\times10^{-2}$ and $1.4\times10^{-2}$, respectively, as shown in the first and second panels of Fig. \ref{fig:RFAP_ET}. To further assess the performance of the two TFA methods, we normalized the maximum amplitude of the waveforms to $5 \times 10^{-21}$. After repeating the same analysis procedure under this normalization, the resulting FAPR distributions are shown in the third and fourth panels of Fig. \ref{fig:RFAP_ET}. With the normalized conditions and a threshold of $\eta = 0.75$, the FAPR values for IMSST and STFT were $1.5\times10^{-2}$ and $9.7\times10^{-4}$, respectively. This result is consistent with the analysis in Fig. \ref{fig:RFAP_LIGO}: both TFA methods yield higher reconstruction event rates when processing amplitude-normalized data, compared to the event rates when the wave source is placed at the same location without normalization. The results in Fig. \ref{fig:RFAP_ET} further indicate that the STFT method has a lower FAPR than the IMSST method, but this typically comes at the cost of low-amplitude signals. Under the same threshold, although the STFT method effectively reduces noise interference, it also leads to a lower reconstruction event rate. It is noteworthy that for both TFA methods, no clear bimodal distribution was observed in the injection signals with higher match score. This suggests that the current analysis methods are insufficient for effectively classifying the burst mechanisms of GW signals based on match score.

\section{Discussion}
\label{sec:dis}


Extracting GWs from CCSN requires more advanced techniques, as their time-frequency characteristics are not as intuitive as those of other common GW sources (such as compact binary coalescences). Therefore, in this study, we simulate data based on the aLIGO and ET detectors and use the IMSST method to reconstruct the GW waveform from CCSN. Additionally, we compare the performance of the IMSST and STFT methods in waveform reconstruction for the ET detector. Using a library of simulated CCSN GW waveforms (shown in Table \ref{tab:waveforms}), we evaluate the performance of these two detectors and TFA methods, quantifying the differences between the reconstructed and injected waveforms through the match score. Finally, we perform an analysis to explore the probability that the extracted GW is a true signal.

We analyzed the maximum distance at which GW signals can be successfully reconstructed using the IMSST method with the aLIGO and ET detectors. By setting a match score threshold of 0.75, we found that the maximum reconstructable distance for the aLIGO detector using IMSST is approximately 37 kpc, while for the ET detector, this distance extends to 317 kpc. The detailed results are presented in Fig. \ref{fig:distance}. To evaluate whether the reconstructed waveforms represent true GW signals rather than noise, we first computed the FAPR for four waveforms at 10 kpc using the aLIGO detector. The FAPR values were all below $2.5 \times 10^{-3}$, confirming the effectiveness and stability of the IMSST method. These results are shown in Fig. \ref{fig:stability}. Next, we extended this analysis to all the waveforms listed in Table \ref{tab:waveforms}. Using the aLIGO detector and the IMSST method, we calculated the FAPR at 10 kpc, which was found to be $2.1 \times 10^{-2}$. To mitigate the impact of amplitude, we normalized the maximum amplitude of the injected GW signals to $5 \times 10^{-21}$, which resulted in a reduced FAPR of $6.2 \times 10^{-3}$. For the ET detector, we computed the FAPR at 100 kpc using the IMSST method, obtaining a value of $1.3 \times 10^{-1}$. After applying the same normalization procedure, the FAPR value was reduced to $1.5 \times 10^{-2}$. The corresponding results for the ET detector are shown in Fig. \ref{fig:RFAP_ET}. In both cases, for the aLIGO and ET detectors, we observed that due to the detectors' response characteristics and the relatively low amplitudes of some GW waveforms, approximately half of the waveforms could not be successfully reconstructed at fixed positions (see the first and third panels of Fig. \ref{fig:RFAP_LIGO}). However, after normalizing the amplitude, only a small fraction of the waveforms failed to be reconstructed successfully (see the second and fourth panels of Fig. \ref{fig:RFAP_LIGO}). Additionally, we compared the performance of the IMSST and STFT TFA methods for waveform reconstruction using the ET detector. We found that the maximum distance at which GW signals could be reconstructed using the IMSST method was 317 kpc, compared to 186 kpc for the STFT method. We also analyzed the FAPR for both methods. When the source distance was fixed at 100 kpc, the FAPR values obtained using the IMSST and STFT methods were $1.3\times10^{-1}$ and $5.6\times10^{-2}$, respectively. After normalizing the maximum amplitude of the GW waveforms to $5 \times 10^{-21}$, the FAPR values for IMSST and STFT were reduced to $1.5\times10^{-2}$ and $9.7\times10^{-4}$, respectively. The corresponding results are shown in Fig. \ref{fig:RFAP_ET}. From the results in Fig. \ref{fig:IM_ST_distance} and Fig. \ref{fig:RFAP_ET}, it is evident that the STFT method is more amplitude-dependent than the IMSST method. While this dependency can effectively reduce the FAPR when reconstructing GW signals, it also leads to a decrease in the event reconstruction rate. Therefore, in practical applications, a trade-off must be made between the two methods to balance the FAPR and the reconstruction event rate.

Compared to the maximum reconstruction distance obtained using the ensemble empirical mode decomposition method \citep{Yuan2024MNRAS}, the results from IMSST and STFT do not show a significant improvement. This is because the time-frequency structure of GW signals generated by CCSN is not yet well understood. As a result, signal reconstruction must be performed over a broad time-frequency space, inevitably introducing substantial noise into the reconstructed signal. In this study, we have only discussed the results using a single detector. Previous studies have explored the use of GW detector networks to search for such events \citep{Klimenko2016PhRvD}, and some have considered the use of space-borne detectors for this purpose \citep{Wu2024PhRvD}. With the continuous advancements in numerical simulation techniques and the ongoing innovation in TFA methods, it is expected that in the near future we will gain a clearer understanding of the time-frequency structure of these GW signals, enabling more accurate reconstruction of signals from greater distances.


\section*{Acknowledgements}

We thank Jade Powell and Man Leong Chan for offering SN waveforms and helpful discussions. This work is supported by the National Key Research and Development Program of China, Grant No. 2021YFC2201901, and the International Partnership Program of the Chinese Academy of Sciences, Grant No.
025GJHZ2023106GC.


\section*{Data Availability}

This theoretical study did not generate any new data. The magnetorotational mechanism CCSN signals were obtained from \url{https://stellarcollapse.org/gwcatalog.html} and \url{https://zenodo.org/record/201145}. The neutrino-driven mechanism CCSN signals were obtained from \url{https://wwwmpa.mpa-garching.mpg.de/ccsnarchive/data/Andresen2019/} and \url{https://www.astro.princeton.edu/~burrows/gw.3d/}.



\bibliographystyle{mnras}
\bibliography{example} 

\begin{thebibliography}{}
\makeatletter
\relax
\def\mn@urlcharsother{\let\do\@makeother \do\$\do\&\do\#\do\^\do\_\do\%\do\~}
\def\mn@doi{\begingroup\mn@urlcharsother \@ifnextchar [ {\mn@doi@}
  {\mn@doi@[]}}
\def\mn@doi@[#1]#2{\def\@tempa{#1}\ifx\@tempa\@empty \href
  {http://dx.doi.org/#2} {doi:#2}\else \href {http://dx.doi.org/#2} {#1}\fi
  \endgroup}
\def\mn@eprint#1#2{\mn@eprint@#1:#2::\@nil}
\def\mn@eprint@arXiv#1{\href {http://arxiv.org/abs/#1} {{\tt arXiv:#1}}}
\def\mn@eprint@dblp#1{\href {http://dblp.uni-trier.de/rec/bibtex/#1.xml}
  {dblp:#1}}
\def\mn@eprint@#1:#2:#3:#4\@nil{\def\@tempa {#1}\def\@tempb {#2}\def\@tempc
  {#3}\ifx \@tempc \@empty \let \@tempc \@tempb \let \@tempb \@tempa \fi \ifx
  \@tempb \@empty \def\@tempb {arXiv}\fi \@ifundefined
  {mn@eprint@\@tempb}{\@tempb:\@tempc}{\expandafter \expandafter \csname
  mn@eprint@\@tempb\endcsname \expandafter{\@tempc}}}

\bibitem[\protect\citeauthoryear{{Abadie} et~al.,}{{Abadie}
  et~al.}{2012}]{Abadie2012PhRvD}
{Abadie} J.,  et~al., 2012, \mn@doi [\prd] {10.1103/PhysRevD.85.122007}, \href
  {https://ui.adsabs.harvard.edu/abs/2012PhRvD..85l2007A} {85, 122007}

\bibitem[\protect\citeauthoryear{{Abbott} et~al.,}{{Abbott}
  et~al.}{2016a}]{Abbott2016PhRvD}
{Abbott} B.~P.,  et~al., 2016a, \mn@doi [\prd] {10.1103/PhysRevD.94.102001},
  \href {https://ui.adsabs.harvard.edu/abs/2016PhRvD..94j2001A} {94, 102001}

\bibitem[\protect\citeauthoryear{{Abbott} et~al.,}{{Abbott}
  et~al.}{2016b}]{Abbott2016PhRvL}
{Abbott} B.~P.,  et~al., 2016b, \mn@doi [\prl]
  {10.1103/PhysRevLett.116.061102}, \href
  {https://ui.adsabs.harvard.edu/abs/2016PhRvL.116f1102A} {116, 061102}

\bibitem[\protect\citeauthoryear{{Abbott} et~al.,}{{Abbott}
  et~al.}{2017a}]{Abbott2017CQGra}
{Abbott} B.~P.,  et~al., 2017a, \mn@doi [Classical and Quantum Gravity]
  {10.1088/1361-6382/aa51f4}, \href
  {https://ui.adsabs.harvard.edu/abs/2017CQGra..34d4001A} {34, 044001}

\bibitem[\protect\citeauthoryear{{Abbott} et~al.,}{{Abbott}
  et~al.}{2017b}]{Abbott2017PhRvD}
{Abbott} B.~P.,  et~al., 2017b, \mn@doi [\prd] {10.1103/PhysRevD.95.042003},
  \href {https://ui.adsabs.harvard.edu/abs/2017PhRvD..95d2003A} {95, 042003}

\bibitem[\protect\citeauthoryear{{Abbott} et~al.,}{{Abbott}
  et~al.}{2019}]{Abbott2019PhRvD}
{Abbott} B.~P.,  et~al., 2019, \mn@doi [\prd] {10.1103/PhysRevD.100.024017},
  \href {https://ui.adsabs.harvard.edu/abs/2019PhRvD.100b4017A} {100, 024017}

\bibitem[\protect\citeauthoryear{{Abbott} et~al.,}{{Abbott}
  et~al.}{2020a}]{Abbott2020LRR}
{Abbott} B.~P.,  et~al., 2020a, \mn@doi [Living Reviews in Relativity]
  {10.1007/s41114-020-00026-9}, \href
  {https://ui.adsabs.harvard.edu/abs/2020LRR....23....3A} {23, 3}

\bibitem[\protect\citeauthoryear{{Abbott} et~al.,}{{Abbott}
  et~al.}{2020b}]{Abbott2020PhRvD}
{Abbott} B.~P.,  et~al., 2020b, \mn@doi [\prd] {10.1103/PhysRevD.101.084002},
  \href {https://ui.adsabs.harvard.edu/abs/2020PhRvD.101h4002A} {101, 084002}

\bibitem[\protect\citeauthoryear{{Abbott} et~al.,}{{Abbott}
  et~al.}{2021}]{Abbott2021PhRvD}
{Abbott} R.,  et~al., 2021, \mn@doi [\prd] {10.1103/PhysRevD.104.122004}, \href
  {https://ui.adsabs.harvard.edu/abs/2021PhRvD.104l2004A} {104, 122004}

\bibitem[\protect\citeauthoryear{{Abbott} et~al.,}{{Abbott}
  et~al.}{2022}]{Abbott2022PTEP}
{Abbott} R.,  et~al., 2022, \mn@doi [Progress of Theoretical and Experimental
  Physics] {10.1093/ptep/ptac073}, \href
  {https://ui.adsabs.harvard.edu/abs/2022PTEP.2022f3F01A} {2022, 063F01}

\bibitem[\protect\citeauthoryear{{Abbott} et~al.,}{{Abbott}
  et~al.}{2023}]{Abbott2021GWTC3}
{Abbott} R.,  et~al., 2023, \mn@doi [Physical Review X]
  {10.1103/PhysRevX.13.041039}, \href
  {https://ui.adsabs.harvard.edu/abs/2023PhRvX..13d1039A} {13, 041039}

\bibitem[\protect\citeauthoryear{{Abdikamalov}, {Gossan}, {DeMaio}  \&
  {Ott}}{{Abdikamalov} et~al.}{2014}]{Abdikamalov2014PhRvD}
{Abdikamalov} E.,  {Gossan} S.,  {DeMaio} A.~M.,   {Ott} C.~D.,  2014, \mn@doi
  [\prd] {10.1103/PhysRevD.90.044001}, \href
  {https://ui.adsabs.harvard.edu/abs/2014PhRvD..90d4001A} {90, 044001}

\bibitem[\protect\citeauthoryear{{Acernese} et~al.,}{{Acernese}
  et~al.}{2015}]{Acernese_2015}
{Acernese} F.,  et~al., 2015, \mn@doi [Classical and Quantum Gravity]
  {10.1088/0264-9381/32/2/024001}, \href
  {https://ui.adsabs.harvard.edu/abs/2015CQGra..32b4001A} {32, 024001}

\bibitem[\protect\citeauthoryear{{Ackley} et~al.,}{{Ackley}
  et~al.}{2020}]{Ackley2020PASA}
{Ackley} K.,  et~al., 2020, \mn@doi [\pasa] {10.1017/pasa.2020.39}, \href
  {https://ui.adsabs.harvard.edu/abs/2020PASA...37...47A} {37, e047}

\bibitem[\protect\citeauthoryear{{Andresen} \& {Finkel}}{{Andresen} \&
  {Finkel}}{2024}]{Andresen2024arXiv}
{Andresen} H.,  {Finkel} B.,  2024, \mn@doi [arXiv e-prints]
  {10.48550/arXiv.2411.12524}, \href
  {https://ui.adsabs.harvard.edu/abs/2024arXiv241112524A} {p. arXiv:2411.12524}

\bibitem[\protect\citeauthoryear{{Andresen}, {M{\"u}ller}, {M{\"u}ller}  \&
  {Janka}}{{Andresen} et~al.}{2017}]{Andresen_mn2017}
{Andresen} H.,  {M{\"u}ller} B.,  {M{\"u}ller} E.,   {Janka} H.~T.,  2017,
  \mn@doi [\mnras] {10.1093/mnras/stx618}, \href
  {https://ui.adsabs.harvard.edu/abs/2017MNRAS.468.2032A} {468, 2032}

\bibitem[\protect\citeauthoryear{{Andresen}, {M{\"u}ller}, {Janka}, {Summa},
  {Gill}  \& {Zanolin}}{{Andresen} et~al.}{2019}]{Andresen2019MNRAS}
{Andresen} H.,  {M{\"u}ller} E.,  {Janka} H.~T.,  {Summa} A.,  {Gill} K.,
  {Zanolin} M.,  2019, \mn@doi [\mnras] {10.1093/mnras/stz990}, \href
  {https://ui.adsabs.harvard.edu/abs/2019MNRAS.486.2238A} {486, 2238}

\bibitem[\protect\citeauthoryear{Auger \& Flandrin}{Auger \&
  Flandrin}{1995}]{auger1995improving}
Auger F.,  Flandrin P.,  1995, IEEE Transactions on signal processing, 43, 1068

\bibitem[\protect\citeauthoryear{Auger, Flandrin, Lin, McLaughlin, Meignen,
  Oberlin  \& Wu}{Auger et~al.}{2013}]{auger2013time}
Auger F.,  Flandrin P.,  Lin Y.-T.,  McLaughlin S.,  Meignen S.,  Oberlin T.,
  Wu H.-T.,  2013, IEEE Signal Processing Magazine, 30, 32

\bibitem[\protect\citeauthoryear{{Baron} \& {Cooperstein}}{{Baron} \&
  {Cooperstein}}{1990}]{Baron_apj1990}
{Baron} E.,  {Cooperstein} J.,  1990, \mn@doi [\apj] {10.1086/168649}, \href
  {https://ui.adsabs.harvard.edu/abs/1990ApJ...353..597B} {353, 597}

\bibitem[\protect\citeauthoryear{{Bethe}}{{Bethe}}{1990}]{Bethe_1990}
{Bethe} H.~A.,  1990, \mn@doi [Reviews of Modern Physics]
  {10.1103/RevModPhys.62.801}, \href
  {https://ui.adsabs.harvard.edu/abs/1990RvMP...62..801B} {62, 801}

\bibitem[\protect\citeauthoryear{{Bethe} \& {Wilson}}{{Bethe} \&
  {Wilson}}{1985}]{Bethe_apj1985}
{Bethe} H.~A.,  {Wilson} J.~R.,  1985, \mn@doi [\apj] {10.1086/163343}, \href
  {https://ui.adsabs.harvard.edu/abs/1985ApJ...295...14B} {295, 14}

\bibitem[\protect\citeauthoryear{{Blondin}, {Mezzacappa}  \&
  {DeMarino}}{{Blondin} et~al.}{2003}]{Blondin_apj2003}
{Blondin} J.~M.,  {Mezzacappa} A.,   {DeMarino} C.,  2003, \mn@doi [\apj]
  {10.1086/345812}, \href
  {https://ui.adsabs.harvard.edu/abs/2003ApJ...584..971B} {584, 971}

\bibitem[\protect\citeauthoryear{{Burrows} \& {Vartanyan}}{{Burrows} \&
  {Vartanyan}}{2021}]{Burrows2021Natur}
{Burrows} A.,  {Vartanyan} D.,  2021, \mn@doi [\nat]
  {10.1038/s41586-020-03059-w}, \href
  {https://ui.adsabs.harvard.edu/abs/2021Natur.589...29B} {589, 29}

\bibitem[\protect\citeauthoryear{{Chan}, {Heng}  \& {Messenger}}{{Chan}
  et~al.}{2020}]{Chan2020PhRvD}
{Chan} M.~L.,  {Heng} I.~S.,   {Messenger} C.,  2020, \mn@doi [\prd]
  {10.1103/PhysRevD.102.043022}, \href
  {https://ui.adsabs.harvard.edu/abs/2020PhRvD.102d3022C} {102, 043022}

\bibitem[\protect\citeauthoryear{Daubechies, Lu  \& Wu}{Daubechies
  et~al.}{2011}]{daubechies2011synchrosqueezed}
Daubechies I.,  Lu J.,   Wu H.-T.,  2011, Applied and computational harmonic
  analysis, 30, 243

\bibitem[\protect\citeauthoryear{{Dimmelmeier}, {Ott}, {Marek}  \&
  {Janka}}{{Dimmelmeier} et~al.}{2008}]{Dimmelmeier2008PhRvD}
{Dimmelmeier} H.,  {Ott} C.~D.,  {Marek} A.,   {Janka} H.~T.,  2008, \mn@doi
  [\prd] {10.1103/PhysRevD.78.064056}, \href
  {https://ui.adsabs.harvard.edu/abs/2008PhRvD..78f4056D} {78, 064056}

\bibitem[\protect\citeauthoryear{Gabor}{Gabor}{1946}]{gabor1946theory}
Gabor D.,  1946, Journal of the Institution of Electrical Engineers-part III:
  radio and communication engineering, 93, 429

\bibitem[\protect\citeauthoryear{{Gossan}, {Sutton}, {Stuver}, {Zanolin},
  {Gill}  \& {Ott}}{{Gossan} et~al.}{2016}]{Gossan2016PhRvD}
{Gossan} S.~E.,  {Sutton} P.,  {Stuver} A.,  {Zanolin} M.,  {Gill} K.,   {Ott}
  C.~D.,  2016, \mn@doi [\prd] {10.1103/PhysRevD.93.042002}, \href
  {https://ui.adsabs.harvard.edu/abs/2016PhRvD..93d2002G} {93, 042002}

\bibitem[\protect\citeauthoryear{{Hayama}, {Kuroda}, {Kotake}  \&
  {Takiwaki}}{{Hayama} et~al.}{2015}]{Hayama_prd2015}
{Hayama} K.,  {Kuroda} T.,  {Kotake} K.,   {Takiwaki} T.,  2015, \mn@doi [\prd]
  {10.1103/PhysRevD.92.122001}, \href
  {https://ui.adsabs.harvard.edu/abs/2015PhRvD..92l2001H} {92, 122001}

\bibitem[\protect\citeauthoryear{{Heng}}{{Heng}}{2009}]{Heng2009CQG}
{Heng} I.~S.,  2009, \mn@doi [Classical and Quantum Gravity]
  {10.1088/0264-9381/26/10/105005}, \href
  {https://ui.adsabs.harvard.edu/abs/2009CQGra..26j5005H} {26, 105005}

\bibitem[\protect\citeauthoryear{{Hild} et~al.,}{{Hild}
  et~al.}{2011}]{Hild2011CQGra}
{Hild} S.,  et~al., 2011, \mn@doi [Classical and Quantum Gravity]
  {10.1088/0264-9381/28/9/094013}, \href
  {https://ui.adsabs.harvard.edu/abs/2011CQGra..28i4013H} {28, 094013}

\bibitem[\protect\citeauthoryear{{Hu}, {Lin}, {Pan}, {Li}, {Yen}, {Kong}  \&
  {Hui}}{{Hu} et~al.}{2022}]{Hu2022ApJ}
{Hu} C.-P.,  {Lin} L. C.-C.,  {Pan} K.-C.,  {Li} K.-L.,  {Yen} C.-C.,  {Kong}
  A. K.~H.,   {Hui} C.~Y.,  2022, \mn@doi [\apj] {10.3847/1538-4357/ac8165},
  \href {https://ui.adsabs.harvard.edu/abs/2022ApJ...935..127H} {935, 127}

\bibitem[\protect\citeauthoryear{{Janka}}{{Janka}}{2012}]{Janka_2012}
{Janka} H.-T.,  2012, \mn@doi [Annual Review of Nuclear and Particle Science]
  {10.1146/annurev-nucl-102711-094901}, \href
  {https://ui.adsabs.harvard.edu/abs/2012ARNPS..62..407J} {62, 407}

\bibitem[\protect\citeauthoryear{{Janka}, {Langanke}, {Marek},
  {Mart{\'\i}nez-Pinedo}  \& {M{\"u}ller}}{{Janka} et~al.}{2007}]{Janka2007PhR}
{Janka} H.~T.,  {Langanke} K.,  {Marek} A.,  {Mart{\'\i}nez-Pinedo} G.,
  {M{\"u}ller} B.,  2007, \mn@doi [\physrep] {10.1016/j.physrep.2007.02.002},
  \href {https://ui.adsabs.harvard.edu/abs/2007PhR...442...38J} {442, 38}

\bibitem[\protect\citeauthoryear{{Janka}, {Melson}  \& {Summa}}{{Janka}
  et~al.}{2016}]{Janka2016ARNPS}
{Janka} H.-T.,  {Melson} T.,   {Summa} A.,  2016, \mn@doi [Annual Review of
  Nuclear and Particle Science] {10.1146/annurev-nucl-102115-044747}, \href
  {https://ui.adsabs.harvard.edu/abs/2016ARNPS..66..341J} {66, 341}

\bibitem[\protect\citeauthoryear{{Klimenko} et~al.,}{{Klimenko}
  et~al.}{2016}]{Klimenko2016PhRvD}
{Klimenko} S.,  et~al., 2016, \mn@doi [\prd] {10.1103/PhysRevD.93.042004},
  \href {https://ui.adsabs.harvard.edu/abs/2016PhRvD..93d2004K} {93, 042004}

\bibitem[\protect\citeauthoryear{{Kotake}, {Sato}  \& {Takahashi}}{{Kotake}
  et~al.}{2006}]{Kotake2006RPPh}
{Kotake} K.,  {Sato} K.,   {Takahashi} K.,  2006, \mn@doi [Reports on Progress
  in Physics] {10.1088/0034-4885/69/4/R03}, \href
  {https://ui.adsabs.harvard.edu/abs/2006RPPh...69..971K} {69, 971}

\bibitem[\protect\citeauthoryear{{Kotake}, {Sumiyoshi}, {Yamada}, {Takiwaki},
  {Kuroda}, {Suwa}  \& {Nagakura}}{{Kotake} et~al.}{2012}]{Kotake_2012PTEP}
{Kotake} K.,  {Sumiyoshi} K.,  {Yamada} S.,  {Takiwaki} T.,  {Kuroda} T.,
  {Suwa} Y.,   {Nagakura} H.,  2012, \mn@doi [Progress of Theoretical and
  Experimental Physics] {10.1093/ptep/pts009}, \href
  {https://ui.adsabs.harvard.edu/abs/2012PTEP.2012aA301K} {2012, 01A301}

\bibitem[\protect\citeauthoryear{{Kuroda}, {Kotake}  \& {Takiwaki}}{{Kuroda}
  et~al.}{2016}]{Kuroda2016ApJL}
{Kuroda} T.,  {Kotake} K.,   {Takiwaki} T.,  2016, \mn@doi [\apjl]
  {10.3847/2041-8205/829/1/L14}, \href
  {https://ui.adsabs.harvard.edu/abs/2016ApJ...829L..14K} {829, L14}

\bibitem[\protect\citeauthoryear{{Kuroda}, {Kotake}, {Hayama}  \&
  {Takiwaki}}{{Kuroda} et~al.}{2017}]{Kuroda2017ApJ}
{Kuroda} T.,  {Kotake} K.,  {Hayama} K.,   {Takiwaki} T.,  2017, \mn@doi [\apj]
  {10.3847/1538-4357/aa988d}, \href
  {https://ui.adsabs.harvard.edu/abs/2017ApJ...851...62K} {851, 62}

\bibitem[\protect\citeauthoryear{{LIGO Scientific Collaboration} et~al.,}{{LIGO
  Scientific Collaboration} et~al.}{2015}]{LIGO_2015CQGra}
{LIGO Scientific Collaboration} et~al., 2015, \mn@doi [Classical and Quantum
  Gravity] {10.1088/0264-9381/32/7/074001}, \href
  {https://ui.adsabs.harvard.edu/abs/2015CQGra..32g4001L} {32, 074001}

\bibitem[\protect\citeauthoryear{{Maggiore} et~al.,}{{Maggiore}
  et~al.}{2020}]{Maggiore2020JCAP}
{Maggiore} M.,  et~al., 2020, \mn@doi [\jcap] {10.1088/1475-7516/2020/03/050},
  \href {https://ui.adsabs.harvard.edu/abs/2020JCAP...03..050M} {2020, 050}

\bibitem[\protect\citeauthoryear{{McIver}}{{McIver}}{2015}]{McIver2015}
{McIver} J.~L.,  2015, PhD thesis, University of Massachusetts Amherst,
  \mn@doi{10.7275/7537749.0}

\bibitem[\protect\citeauthoryear{{Mezzacappa} \& {Zanolin}}{{Mezzacappa} \&
  {Zanolin}}{2024}]{Mezzacappa2024arXiv}
{Mezzacappa} A.,  {Zanolin} M.,  2024, \mn@doi [arXiv e-prints]
  {10.48550/arXiv.2401.11635}, \href
  {https://ui.adsabs.harvard.edu/abs/2024arXiv240111635M} {p. arXiv:2401.11635}

\bibitem[\protect\citeauthoryear{{Mezzacappa} et~al.,}{{Mezzacappa}
  et~al.}{2014}]{Mezzacappa_2014}
{Mezzacappa} A.,  et~al., 2014, in {Pogorelov} N.~V.,  {Audit} E.,   {Zank}
  G.~P.,  eds,  Astronomical Society of the Pacific Conference Series Vol. 488,
  8th International Conference of Numerical Modeling of Space Plasma Flows
  (ASTRONUM 2013). p.~102 (\mn@eprint {arXiv} {1405.7075}),
  \mn@doi{10.48550/arXiv.1405.7075}

\bibitem[\protect\citeauthoryear{{Mitra}, {Orel}, {Abylkairov}, {Shukirgaliyev}
   \& {Abdikamalov}}{{Mitra} et~al.}{2024}]{Mitra2024MNRAS}
{Mitra} A.,  {Orel} D.,  {Abylkairov} Y.~S.,  {Shukirgaliyev} B.,
  {Abdikamalov} E.,  2024, \mn@doi [\mnras] {10.1093/mnras/stae714}, \href
  {https://ui.adsabs.harvard.edu/abs/2024MNRAS.529.3582M} {529, 3582}

\bibitem[\protect\citeauthoryear{{Moenchmeyer}, {Schaefer}, {Mueller}  \&
  {Kates}}{{Moenchmeyer} et~al.}{1991}]{Moenchmeyer1991A&A}
{Moenchmeyer} R.,  {Schaefer} G.,  {Mueller} E.,   {Kates} R.~E.,  1991, \aap,
  \href {https://ui.adsabs.harvard.edu/abs/1991A&A...246..417M} {246, 417}

\bibitem[\protect\citeauthoryear{{Morozova}, {Radice}, {Burrows}  \&
  {Vartanyan}}{{Morozova} et~al.}{2018}]{Morozova2018ApJ}
{Morozova} V.,  {Radice} D.,  {Burrows} A.,   {Vartanyan} D.,  2018, \mn@doi
  [\apj] {10.3847/1538-4357/aac5f1}, \href
  {https://ui.adsabs.harvard.edu/abs/2018ApJ...861...10M} {861, 10}

\bibitem[\protect\citeauthoryear{{M{\"u}ller}}{{M{\"u}ller}}{2016}]{Muller2016PASA}
{M{\"u}ller} B.,  2016, \mn@doi [\pasa] {10.1017/pasa.2016.40}, \href
  {https://ui.adsabs.harvard.edu/abs/2016PASA...33...48M} {33, e048}

\bibitem[\protect\citeauthoryear{{M{\"u}ller}, {Janka}, {Marek}, {Hanke},
  {Wongwatha-narat}  \& {M{\"u}ller}}{{M{\"u}ller}
  et~al.}{2011}]{Muller2011hnse}
{M{\"u}ller} B.,  {Janka} H.-T.,  {Marek} A.,  {Hanke} F.,  {Wongwatha-narat}
  A.,   {M{\"u}ller} E.,  2011, in Hamburg Neutrinos from Supernova Explosions.
  p.~14 (\mn@eprint {arXiv} {1112.1913}), \mn@doi{10.48550/arXiv.1112.1913}

\bibitem[\protect\citeauthoryear{{M{\"u}ller}, {Janka}  \&
  {Wongwathanarat}}{{M{\"u}ller} et~al.}{2012}]{Muller2012AA}
{M{\"u}ller} E.,  {Janka} H.~T.,   {Wongwathanarat} A.,  2012, \mn@doi [\aap]
  {10.1051/0004-6361/201117611}, \href
  {https://ui.adsabs.harvard.edu/abs/2012A&A...537A..63M} {537, A63}

\bibitem[\protect\citeauthoryear{{Nuttall} et~al.,}{{Nuttall}
  et~al.}{2015}]{Nuttall2015CQGra}
{Nuttall} L.~K.,  et~al., 2015, \mn@doi [Classical and Quantum Gravity]
  {10.1088/0264-9381/32/24/245005}, \href
  {https://ui.adsabs.harvard.edu/abs/2015CQGra..32x5005N} {32, 245005}

\bibitem[\protect\citeauthoryear{{Ott} et~al.,}{{Ott}
  et~al.}{2013}]{Ott2013ApJ}
{Ott} C.~D.,  et~al., 2013, \mn@doi [\apj] {10.1088/0004-637X/768/2/115}, \href
  {https://ui.adsabs.harvard.edu/abs/2013ApJ...768..115O} {768, 115}

\bibitem[\protect\citeauthoryear{{Owen} \& {Sathyaprakash}}{{Owen} \&
  {Sathyaprakash}}{1999}]{Owen1999PhRvD}
{Owen} B.~J.,  {Sathyaprakash} B.~S.,  1999, \mn@doi [\prd]
  {10.1103/PhysRevD.60.022002}, \href
  {https://ui.adsabs.harvard.edu/abs/1999PhRvD..60b2002O} {60, 022002}

\bibitem[\protect\citeauthoryear{Pons-Llinares, Antonino-Daviu, Riera-Guasp,
  Lee, Kang  \& Yang}{Pons-Llinares et~al.}{2014}]{pons2014advanced}
Pons-Llinares J.,  Antonino-Daviu J.~A.,  Riera-Guasp M.,  Lee S.~B.,  Kang
  T.-j.,   Yang C.,  2014, IEEE Transactions on Industrial Electronics, 62,
  1791

\bibitem[\protect\citeauthoryear{{Powell} \& {M{\"u}ller}}{{Powell} \&
  {M{\"u}ller}}{2019}]{Powell2019MNRAS}
{Powell} J.,  {M{\"u}ller} B.,  2019, \mn@doi [\mnras] {10.1093/mnras/stz1304},
  \href {https://ui.adsabs.harvard.edu/abs/2019MNRAS.487.1178P} {487, 1178}

\bibitem[\protect\citeauthoryear{{Powell} \& {M{\"u}ller}}{{Powell} \&
  {M{\"u}ller}}{2020}]{Powell2020MNRAS}
{Powell} J.,  {M{\"u}ller} B.,  2020, \mn@doi [\mnras]
  {10.1093/mnras/staa1048}, \href
  {https://ui.adsabs.harvard.edu/abs/2020MNRAS.494.4665P} {494, 4665}

\bibitem[\protect\citeauthoryear{{Powell}, {Iess}, {Llorens-Monteagudo},
  {Obergaulinger}, {M{\"u}ller}, {Torres-Forn{\'e}}, {Cuoco}  \&
  {Font}}{{Powell} et~al.}{2024}]{Powell2024PhRvD}
{Powell} J.,  {Iess} A.,  {Llorens-Monteagudo} M.,  {Obergaulinger} M.,
  {M{\"u}ller} B.,  {Torres-Forn{\'e}} A.,  {Cuoco} E.,   {Font} J.~A.,  2024,
  \mn@doi [\prd] {10.1103/PhysRevD.109.063019}, \href
  {https://ui.adsabs.harvard.edu/abs/2024PhRvD.109f3019P} {109, 063019}

\bibitem[\protect\citeauthoryear{{Punturo} et~al.,}{{Punturo}
  et~al.}{2010}]{Punturo_2010CQGra}
{Punturo} M.,  et~al., 2010, \mn@doi [Classical and Quantum Gravity]
  {10.1088/0264-9381/27/19/194002}, \href
  {https://ui.adsabs.harvard.edu/abs/2010CQGra..27s4002P} {27, 194002}

\bibitem[\protect\citeauthoryear{{Radice}, {Morozova}, {Burrows}, {Vartanyan}
  \& {Nagakura}}{{Radice} et~al.}{2019}]{Radice2019ApJL}
{Radice} D.,  {Morozova} V.,  {Burrows} A.,  {Vartanyan} D.,   {Nagakura} H.,
  2019, \mn@doi [\apjl] {10.3847/2041-8213/ab191a}, \href
  {https://ui.adsabs.harvard.edu/abs/2019ApJ...876L...9R} {876, L9}

\bibitem[\protect\citeauthoryear{{Richers}, {Ott}, {Abdikamalov}, {O'Connor}
  \& {Sullivan}}{{Richers} et~al.}{2017}]{Richers2017PhRvD}
{Richers} S.,  {Ott} C.~D.,  {Abdikamalov} E.,  {O'Connor} E.,   {Sullivan} C.,
   2017, \mn@doi [\prd] {10.1103/PhysRevD.95.063019}, \href
  {https://ui.adsabs.harvard.edu/abs/2017PhRvD..95f3019R} {95, 063019}

\bibitem[\protect\citeauthoryear{{R{\"o}ver}, {Bizouard}, {Christensen},
  {Dimmelmeier}, {Heng}  \& {Meyer}}{{R{\"o}ver} et~al.}{2009}]{Rover2009PhRvd}
{R{\"o}ver} C.,  {Bizouard} M.-A.,  {Christensen} N.,  {Dimmelmeier} H.,
  {Heng} I.~S.,   {Meyer} R.,  2009, \mn@doi [\prd]
  {10.1103/PhysRevD.80.102004}, \href
  {https://ui.adsabs.harvard.edu/abs/2009PhRvD..80j2004R} {80, 102004}

\bibitem[\protect\citeauthoryear{{Scheidegger}, {Fischer}, {Whitehouse}  \&
  {Liebend{\"o}rfer}}{{Scheidegger} et~al.}{2008}]{Scheidegger_2008}
{Scheidegger} S.,  {Fischer} T.,  {Whitehouse} S.~C.,   {Liebend{\"o}rfer} M.,
  2008, \mn@doi [\aap] {10.1051/0004-6361:20078577}, \href
  {https://ui.adsabs.harvard.edu/abs/2008A&A...490..231S} {490, 231}

\bibitem[\protect\citeauthoryear{{Suvorova}, {Powell}  \& {Melatos}}{{Suvorova}
  et~al.}{2019}]{Suvorova2019PhRvD}
{Suvorova} S.,  {Powell} J.,   {Melatos} A.,  2019, \mn@doi [\prd]
  {10.1103/PhysRevD.99.123012}, \href
  {https://ui.adsabs.harvard.edu/abs/2019PhRvD..99l3012S} {99, 123012}

\bibitem[\protect\citeauthoryear{{Szczepa{\'n}czyk} et~al.,}{{Szczepa{\'n}czyk}
  et~al.}{2021}]{Szczepanczyk2021PhRvD}
{Szczepa{\'n}czyk} M.~J.,  et~al., 2021, \mn@doi [\prd]
  {10.1103/PhysRevD.104.102002}, \href
  {https://ui.adsabs.harvard.edu/abs/2021PhRvD.104j2002S} {104, 102002}

\bibitem[\protect\citeauthoryear{{The LIGO Scientific Collaboration}
  et~al.,}{{The LIGO Scientific Collaboration} et~al.}{2024}]{LIGO2024arXiv}
{The LIGO Scientific Collaboration} et~al., 2024, \mn@doi [arXiv e-prints]
  {10.48550/arXiv.2410.16565}, \href
  {https://ui.adsabs.harvard.edu/abs/2024arXiv241016565T} {p. arXiv:2410.16565}

\bibitem[\protect\citeauthoryear{Vartanyan, Burrows, Wang, Coleman  \&
  White}{Vartanyan et~al.}{2023}]{Vartanyan2023PhysRevD}
Vartanyan D.,  Burrows A.,  Wang T.,  Coleman M. S.~B.,   White C.~J.,  2023,
  \mn@doi [Phys. Rev. D] {10.1103/PhysRevD.107.103015}, 107, 103015

\bibitem[\protect\citeauthoryear{Vedre{\~n}o-Santos, Riera-Guasp, Henao,
  Pineda-S{\'a}nchez  \& Puche-Panadero}{Vedre{\~n}o-Santos
  et~al.}{2013}]{vedreno2013diagnosis}
Vedre{\~n}o-Santos F.,  Riera-Guasp M.,  Henao H.,  Pineda-S{\'a}nchez M.,
  Puche-Panadero R.,  2013, IEEE Transactions on Industrial Electronics, 61,
  4947

\bibitem[\protect\citeauthoryear{{Wang} \& {Burrows}}{{Wang} \&
  {Burrows}}{2023}]{Wang2023ApJ}
{Wang} T.,  {Burrows} A.,  2023, \mn@doi [\apj] {10.3847/1538-4357/ace7b2},
  \href {https://ui.adsabs.harvard.edu/abs/2023ApJ...954..114W} {954, 114}

\bibitem[\protect\citeauthoryear{Wang, Chen, Cai, Chen, Li  \& He}{Wang
  et~al.}{2013a}]{wang2013matching}
Wang S.,  Chen X.,  Cai G.,  Chen B.,  Li X.,   He Z.,  2013a, IEEE
  Transactions on Signal Processing, 62, 69

\bibitem[\protect\citeauthoryear{Wang, Chen, Li, Li  \& He}{Wang
  et~al.}{2013b}]{wang2013matchfault}
Wang S.,  Chen X.,  Li G.,  Li X.,   He Z.,  2013b, IEEE Transactions on
  Instrumentation and Measurement, 63, 1372

\bibitem[\protect\citeauthoryear{{Wu}, {Hu}, {Fan}  \& {Heng}}{{Wu}
  et~al.}{2024}]{Wu2024PhRvD}
{Wu} Z.,  {Hu} Y.-M.,  {Fan} H.-M.,   {Heng} I.~S.,  2024, \mn@doi [\prd]
  {10.1103/PhysRevD.109.103004}, \href
  {https://ui.adsabs.harvard.edu/abs/2024PhRvD.109j3004W} {109, 103004}

\bibitem[\protect\citeauthoryear{{Yakunin} et~al.,}{{Yakunin}
  et~al.}{2017a}]{Yakunin_2017}
{Yakunin} K.~N.,  et~al., 2017a, \mn@doi [arXiv e-prints]
  {10.48550/arXiv.1701.07325}, \href
  {https://ui.adsabs.harvard.edu/abs/2017arXiv170107325Y} {p. arXiv:1701.07325}

\bibitem[\protect\citeauthoryear{{Yakunin} et~al.,}{{Yakunin}
  et~al.}{2017b}]{Yakunin2017arXiv}
{Yakunin} K.~N.,  et~al., 2017b, \mn@doi [arXiv e-prints]
  {10.48550/arXiv.1701.07325}, \href
  {https://ui.adsabs.harvard.edu/abs/2017arXiv170107325Y} {p. arXiv:1701.07325}

\bibitem[\protect\citeauthoryear{Yang, Peng, Meng  \& Zhang}{Yang
  et~al.}{2011}]{yang2011spline}
Yang Y.,  Peng Z.,  Meng G.,   Zhang W.,  2011, IEEE Transactions on Industrial
  Electronics, 59, 1612

\bibitem[\protect\citeauthoryear{Yang, Zhang, Peng  \& Meng}{Yang
  et~al.}{2012}]{yang2012multicomponent}
Yang Y.,  Zhang W.,  Peng Z.,   Meng G.,  2012, IEEE Transactions on Industrial
  Electronics, 60, 3948

\bibitem[\protect\citeauthoryear{Yu}{Yu}{2020}]{Yu2020JSV}
Yu G.,  2020, \mn@doi [Journal of Sound and Vibration] {10.33232/001c.117147},
  492, 115813

\bibitem[\protect\citeauthoryear{Yu, Wang  \& Zhao}{Yu
  et~al.}{2018}]{Yu2018MSST}
Yu G.,  Wang Z.,   Zhao P.,  2018, IEEE Transactions on Industrial Electronics,
  66, 5441

\bibitem[\protect\citeauthoryear{{Yuan}, {Fan}, {L{\"u}}, {Sun}  \&
  {Lin}}{{Yuan} et~al.}{2024}]{Yuan2024MNRAS}
{Yuan} Y.,  {Fan} X.-L.,  {L{\"u}} H.-J.,  {Sun} Y.-Y.,   {Lin} K.,  2024,
  \mn@doi [\mnras] {10.1093/mnras/stae604}, \href
  {https://ui.adsabs.harvard.edu/abs/2024MNRAS.529.3235Y} {529, 3235}

\makeatother
\end{thebibliography}








\label{lastpage}
\end{document}